\documentclass[fleqn,usenatbib]{mnras}

\usepackage[T1]{fontenc}

\DeclareRobustCommand{\VAN}[3]{#2}
\let\VANthebibliography\thebibliography
\def\thebibliography{\DeclareRobustCommand{\VAN}[3]{##3}\VANthebibliography}

\usepackage[utf8]{inputenc}

\usepackage{lmodern}
\usepackage{fixcmex}  

\usepackage{microtype}

\usepackage{graphicx}
\usepackage{tikz}

\usepackage{relsize}

\usepackage{scalerel}

\usepackage{calc}

\usepackage{amsmath}
\usepackage{amssymb}
\usepackage{mathtools}

\usepackage{commath}

\usepackage{slashed}

\usepackage{siunitx}

\usepackage{textalpha}

\usepackage{cleveref}

\allowdisplaybreaks

\crefname{equation}{}{}                 
\renewcommand{\eqref}[1]{\cref{#1}}

\newcommand{\PI}{\text{\textpi}}

\newcommand{\cbrt}[1]{\sqrt[3]{#1}}

\renewcommand{\vec}[1]{\mathbfit{#1}}

\DeclareMathOperator{\erf}{erf}
\DeclareMathOperator{\erfc}{erfc}
\DeclareMathOperator{\sinc}{sinc}
\DeclarePairedDelimiter\ceil{\lceil}{\rceil}
\DeclarePairedDelimiter\floor{\lfloor}{\rfloor}

\DeclareMathOperator*{\bigast}{\scalerel*{\ast}{\sum}}

\newcommand{\approptoinn}[2]{\mathrel{\vcenter{\offinterlineskip\halign{\hfil$##$\cr#1\propto\cr\noalign{\kern2pt}#1\sim\cr\noalign{\kern-2pt}}}}}
\newcommand{\appropto}{\mathpalette\approptoinn\relax}

\makeatletter

\renewcommand{\d}[1][]{\dif^{\,#1}\!}
\makeatother

\newcommand{\deltarho}{\text{\textdelta}\rho}

\newcommand{\pshape}{
 {\mathchoice
  {\tikz \shade [inner color=transparent!100, outer color=transparent!0] circle (0.8ex);}
  {\tikz \shade [inner color=transparent!100, outer color=transparent!0] circle (0.8ex);}
  {\tikz \shade [inner color=transparent!100, outer color=transparent!0] circle (0.6ex);}
  {\tikz \shade [inner color=transparent!100, outer color=transparent!0] circle (0.45ex);}
 }
}

\DeclareFontFamily{U}{mathb}{\hyphenchar\font45}
\DeclareFontShape{U}{mathb}{m}{n}{
      <5> <6> <7> <8> <9> <10> gen * mathb
      <10.95> mathb10 <12> <14.4> <17.28> <20.74> <24.88> mathb12
      }{}
\DeclareSymbolFont{mathb}{U}{mathb}{m}{n}
\DeclareFontSubstitution{U}{mathb}{m}{n}
\DeclareFontFamily{U}{mathx}{\hyphenchar\font45}
\DeclareFontShape{U}{mathx}{m}{n}{
      <5> <6> <7> <8> <9> <10>
      <10.95> <12> <14.4> <17.28> <20.74> <24.88>
      mathx10
      }{}
\DeclareSymbolFont{mathx}{U}{mathx}{m}{n}
\DeclareFontSubstitution{U}{mathx}{m}{n}
\DeclareMathDelimiter{\thickvert}{0}{mathb}{"7E}{mathx}{"1F}

\newcommand{\CONCEPT}{\textsc{co\textsl{n}cept}}
\newcommand{\CONCEPTONE}{\textsc{co\textsl{n}cept}\,\textscale{.77}{{1.0}}}
\makeatletter
\DeclareRobustCommand*\pac[1]{%
   \if b\expandafter\@car\f@series\@nil
     \textscale{.77}{{#1}}%
   \else
     {\scshape #1}%
   \fi}
\makeatother
\newcommand{\CONCEPTHEADING}{\pac{CO\textsl{N}CEPT}}
\newcommand{\CONCEPTONEHEADING}{\pac{CO\textsl{N}CEPT\,1.0}}
\newcommand{\GADGET}{\textsc{gadget}}
\newcommand{\GADGETHEADING}{\pac{GADGET}}
\newcommand{\GADGETTWO}{\textsc{gadget}\mbox{-}\textscale{.77}{{2}}}
\newcommand{\GADGETTWOHEADING}{\pac{GADGET\mbox{-}2}}

\newcommand{\GADGETFOUR}{\textsc{gadget}\mbox{-}\textscale{.77}{{4}}}
\newcommand{\GADGETFOURHEADING}{\pac{GADGET\mbox{-}4}}
\newcommand{\GADGETTWOFOUR}{\textsc{gadget}\mbox{-}\textscale{.77}{{2/4}}}
\newcommand{\CLASS}{\textsc{class}}
\newcommand{\COSIRA}{\textsc{cosira}}
\newcommand{\PKDGRAVTHREE}{\textsc{pkdgrav}\textscale{.77}{{3}}}
\newcommand{\CUBEPTHREEM}{\textsc{cubep}\textsuperscript{\textscale{.77}{{3}}\hspace{-.06em}}\textsc{m}}
\newcommand{\CUBE}{\textsc{cube}}

\newcommand{\PTHREEM}{P\textsuperscript{3\hspace{-.06em}}M}
\newcommand{\PTHREEMPAREN}{P\textsuperscript{(3)\hspace{-.06em}}M}
\newcommand{\PTHREEMIT}{P\textsuperscript{\hspace{.08em}3\hspace{-.224em}}M}
\newcommand{\CPP}{C\nolinebreak[4]\hspace{-.05em}\raisebox{.4ex}{\relsize{-3}{\textbf{++}}}}

\newcommand{\textplus}{\raisebox{.4ex}{\relsize{-3}{\textbf{+}}}}

\title[The \CONCEPTONE{} simulation code]{The cosmological simulation code \CONCEPTONEHEADING{}}

\author[Dakin, Hannestad \& Tram]{
Jeppe Dakin$^{1}$\thanks{E-mail: dakin@phys.au.dk},
Steen Hannestad$^{1}$\thanks{E-mail: sth@phys.au.dk},
Thomas Tram$^{1}$\thanks{E-mail: thomas.tram@phys.au.dk}
\\
$^{1}$Department of Physics and Astronomy, Aarhus University, Ny Munkegade 120, DK-8000 Aarhus C, Denmark
\phantom{\raisebox{1.7\baselineskip}{.}}  
}

\date{\today}

\pubyear{2021}

\begin{document}
\label{firstpage}
\pagerange{\pageref{firstpage}--\pageref{lastpage}}
\maketitle

\begin{abstract}
We present version 1.0 of the cosmological simulation code \CONCEPT{}, designed for simulations of large-scale structure formation.
\CONCEPTONE{} contains a \PTHREEM{} gravity solver, with the short-range part implemented using a novel (sub)tiling strategy, coupled with individual and adaptive particle time-stepping. A primary objective of \CONCEPT{} is ease of use. To this end, it has built-in initial condition generation and can produce output in the form of snapshots, power spectra and direct visualisations. \CONCEPT{} is the first massively parallel cosmological simulation code written in Python. Despite of this, excellent performance is obtained, even comparing favourably to other codes such as \GADGET{} at similar precision, in the case of low to moderate clustering. By means of power spectrum comparisons we find extraordinary good agreement between \CONCEPTONE{} and \GADGET{}. At large and intermediate scales the codes agree to well below the per mille level, while the agreement at the smallest scales probed ($k \sim 13\, h/\si{Mpc}$) is of the order of $\SI{1}{\percent}$. The \CONCEPT{} code is openly released and comes with a robust installation script as well as thorough documentation.
\end{abstract}



\section{Introduction}\label{sec:intro}
Measurements of inhomogeneities in our Universe have been performed over a vast range of scales, spanning sub-galactic scales all the way to the current horizon. On large scales and at early times the amplitude of density fluctuations is small enough that it can be treated accurately using perturbation theory. However, on smaller scales and at later times this is no longer the case, and structure formation must be evolved through simulation.

The dominant clustering component is cold dark matter which is well-described by a collisionless fluid with negligible thermal velocity dispersion. This means that the full 6D phase space distribution can be collapsed into 3D position space, which can be followed in time. By far the most efficient way of doing this is to use $N$-body simulations in which the underlying fluid is described by a large number $N$ of discrete particles, each following the appropriate equations of motion. This method has the advantage of being inherently Lagrangian --- regions of high density will automatically correspond to regions of high $N$-body particle count, unlike e.g.\ solving the fluid equations using a static Eulerian grid.

Such simulations of cosmic structure formation have a long history, going back to the pioneering work of \citet{Hoerner1960} who proposed to study stellar clusters using $N$-body methods.
The first papers on $N$-body methods used direct summation to find the individual forces on particles. However, this approach quickly becomes prohibitively expensive for large $N$, given that it is an $\mathcal{O}(N^2)$ problem.

In order to make the problem tractable, a number of numerical schemes have been developed over the years, including tree codes \citep{Barnes:1986nb} and particle-mesh (PM) codes \citep{HockneyEastwood}. Tree methods work by first grouping the particles into nodes in a hierarchical tree structure, which is then `walked' to some sufficient depth relative to a given particle in order to provide an approximate but cheap estimate of the gravitational force from several other particles at once. In PM codes a density field on a grid is constructed through interpolation of the particles, which is then transformed to the gravitational potential, typically using fast Fourier techniques. The PM method is much faster than direct summation for large $N$, scaling as $\mathcal{O}(N \log N)$. Though tree codes have a similar scaling $\mathcal{O}(N \log N)$, they are not as fast as PM codes. However, as pure PM codes are restricted by the finite size of the grid cells, this limits their resolution to scales a few times larger than this size.

The shortcomings of the PM method can be mended by augmenting it with direct summation of particle forces over short distances. This method was first described in \citet*{Hockney74} and applied to a cosmological setting by~\citet{Efstathiou1981}. It is known as PP-PM, or \PTHREEM{} \citep{HockneyEastwood,Bertschinger98,HarnoisDeraps:2012vd}.

While \PTHREEM{} codes work extremely well for large-scale cosmological simulations in which clustering is moderate, the non-hierarchical nature of the short-range force becomes an issue when the matter distribution becomes very uneven, e.g.\ for a close-up simulation of a single galaxy formation. This serious problem can be circumvented by using either a tree decomposition of the short-range force (as in the TreePM method of \GADGET{}~\citep{gadget2}), or by applying adaptive mesh refinement to the PM grid~\citep{Couchman91}.

This paper is about release 1.0 of the \CONCEPT{} code\footnote{The \CONCEPT{} code itself along with documentation is openly released at \href{https://github.com/jmd-dk/concept}{github.com/jmd-dk/concept}\,.}, a massively parallel simulation code for cosmological structure formation. The main goal of any such code is to track the non-linear evolution of matter, which \CONCEPT{} achieves through $N$-body techniques, i.e.\ by describing matter as a set of Lagrangian particles. Additionally, \CONCEPT{} allows for any species to be modelled as a (linear or non-linear) fluid, with quantities like energy density, momentum density and pressure being evolved on a spatially fixed, Eulerian grid. This allows for non-standard simulations, such as ones including non-linearly evolved massive neutrinos \citep{nuconcept} and ones fully consistent with general relativistic perturbation theory \citep{concept_linnu,concept_de}. Decaying dark matter scenarios are supported as well \citep*{concept_dcdm}. These more exotic aspects of \CONCEPT{} date back to previous releases and will not be described in detail in this paper.

The main feature new to the 1.0 release of \CONCEPT{} is that of explicit short-range gravitational forces. Previously, the only feasible\footnote{An inefficient implementation of \PTHREEM{} has in fact been available for years. The basic PP method was (and still is) available as well, though due to its $\mathcal{O}(N^2)$ scaling this is intended only for internal testing.} gravitational method available was that of PM, leaving gravity badly resolved at small scales. In \CONCEPTONE{} the extremely fast PM method is retained, though the default gravitational solver is now that of \PTHREEM{}, i.e.\ long-ranged PM augmented with short-ranged direct summation. This newly added short-range force is implemented using an efficient and novel scheme, based on what we call tiles and subtiles. The increased spatial resolution resulting from the added short-range forces calls for a corresponding increase in the temporal resolution, in regions of high clustering. Thus, \CONCEPTONE{} further comes with a new, individual and adaptive particle time-stepping scheme.

The main goal of this paper is threefold: 1) To describe the numerical methods employed by \CONCEPTONE{}, 2) to demonstrate the validity of the code by comparing its results to those of other simulation codes, 3) to measure the code performance in terms of scaling\footnote{Currently the largest simulation performed with \CONCEPTONE{} has $\sim 65$ billion particles. The scaling tests presented in this paper are concerned with more typical simulation sizes.} behaviour as well as absolute comparison to other codes. For the code comparisons, we use the well-known \GADGETTWO{} code~\citep{gadget2} as well as its newer incarnation \GADGETFOUR{}~\citep{gadget4}.

This paper is structured as follows: In section~\ref{sec:numerical_methods} we describe the numerical methods built into \CONCEPTONE{}, with a focus on gravity and time-stepping. Section~\ref{sec:validation_comparison} then goes on to validate the code results, while code performance is explored in section~\ref{sec:performance}. Finally, section~\ref{sec:conclusion} provides a summary and a discussion about the usefulness of the code as it currently stands, as well as what might be implemented in the future in order to enhance both its capabilities and performance. In addition, other features and non-standard software aspects of \CONCEPTONE{} are briefly introduced in appendix~\ref{sec:other}.

\section{Numerical methods}\label{sec:numerical_methods}
This section describes the main numerical methods and implementations used in \CONCEPTONE{}, responsible for the gravitational interaction between matter particles and their resulting temporal evolution.

The basic setup of \CONCEPT{} is that of a cubic, toroidal periodic box of constant comoving side length $L_{\text{box}}$, containing $N$ matter particles of equal mass $m$, each having a comoving position $\vec{x}_i(t)$ and canonical momentum $\vec{q}_i(t)$, evolving under self-gravity in an expanding background, captured by the cosmological scale factor $a(t)$, with $t$ being cosmic time. The code is parallelised using the Message Passing Interface (MPI), with the box divided into equally shaped cuboidal domains --- one per process --- which in turn are mapped one-to-one to physical CPU cores.

The equations of motion for the particles are fully written out in section~\ref{subsec:timestepping}. Before that, section~\ref{subsec:gravity} sets out to find the comoving gravitational force $\vec{f}_i$, the only force considered; $\partial_t \vec{q}_i \equiv \vec{f}_i/a$.

\subsection{Gravity}\label{subsec:gravity}
This subsection develops the gravitational solvers available in \CONCEPTONE{}, starting with the PP and PM method and culminating in the \PTHREEM{} method. While the \CONCEPTONE{} implementations of PP and PM does not deviate much from standard procedures, the \PTHREEM{} implementation is novel.

\subsubsection{PP gravity}
The particle-particle (PP) method solves gravity via direct summation over pairwise interactions. This direct approach makes the PP method essentially exact, but comes at the cost of $\mathcal{O}(N^2)$ complexity, drastically limiting its usability. Regardless, the PP method is worth studying in detail as it introduces many aspects used for the superior \PTHREEM{} method.

\paragraph*{From particles to fields}
For a set of $N$ point particles in infinite space one could simply use Newton's law of universal gravitation. As we seek more flexibility we shall instead think in terms of the peculiar potential $\varphi$, defined through the Poisson equation \citep{Peebles1980}
\begin{equation}
	\nabla^2\varphi(\vec{x}) = 4\PI Ga^2 \deltarho(\vec{x}) \,, \label{eq:Poisson}
\end{equation}
where $G$ is the gravitational constant and the Laplacian is to be taken with respect to comoving space $\vec{x} \equiv \vec{r}/a(t)$, $\vec{r}$ being physical space. The physical density contrast field $\deltarho(\vec{x})$ is constructed from the particles by assigning them a localised shape $S(\vec{x})$, so that
\begin{equation}
	\deltarho(\vec{x}) = \frac{m}{a^3} \sum_{\vec{n}\in\mathbb{Z}^3} \Biggl\{ -\frac{N}{L_{\text{box}}^{3}} + \sum_{j=1}^N   S(\vec{x} - \vec{x}_{j\vec{n}}) \Biggr\}\,, \label{eq:density_contrast}
\end{equation}
with $\vec{x}_{j\vec{n}} \equiv \vec{x}_j + L_{\text{box}}\vec{n}$ and the periodicity of the box implemented by the sum over all integer triplets $\vec{n}$. We shall refer to the infinitely many particles at $\vec{x}_{j\vec{n}}$, $\vec{n}\in\mathbb{Z}^3$, as particle \emph{images}. The subtraction of $N$ times the reciprocal box volume in \eqref{eq:density_contrast} ensures that $\deltarho(\vec{x})$ averages to zero, assuming the shape $S$ to be normalised to unity.

For point particles, $S(\vec{x}) \rightarrow S_\delta(\vec{x}) \equiv \delta^3(\vec{x})$, $\delta^3$ being the three-dimensional Dirac delta function. Given a shape, \eqref{eq:Poisson} and \eqref{eq:density_contrast} can be solved for the potential;
\begin{equation}
	\varphi(\vec{x}) = -\frac{Gm}{a} \sum_{j=1}^N  \sum_{\vec{n}\in\mathbb{Z}^3} |\vec{x} - \vec{x}_{j\vec{n}}|_{\pshape}^{-1} \,, \label{eq:potential}
\end{equation}
where we have introduced the generalised reciprocal distance $|\vec{x}|^{-1}_{\pshape}$, the subscript denoting an arbitrary shape. For the choice of point particles $S_\delta$, we simply have $|\vec{x}|^{-1}_{\pshape} \rightarrow |\vec{x}|^{-1}_\delta = |\vec{x}|^{-1}$. The comoving force on particle $i$, $\vec{f}_i = -am\eval[0]{\nabla \varphi(\vec{x})}_{\vec{x}=\vec{x}_i}$, is then
\begin{equation}
	\vec{f}_i = -Gm^2 \sum_{\substack{j=1 \\ j \neq i}}^N \sum_{\vec{n}\in\mathbb{Z}^3} |\vec{x}_{ij\vec{n}}|^{-3}_{\pshape} \vec{x}_{ij\vec{n}} \,, \label{eq:force}
\end{equation}
where $\vec{x}_{ij\vec{n}} \equiv \vec{x}_i - \vec{x}_{j\vec{n}} = \vec{x}_i - (\vec{x}_j + L_{\text{box}}\vec{n})$ and the divergence at $j = i$ has been removed. For point particles $|\vec{x}|^{-3}_\delta = (|\vec{x}|^{-1}_\delta)^3 = |\vec{x}|^{-3}$.

\paragraph*{Softening}
As the $N$ tracer particles are meant to represent an underlying continuous density field, it is desirable to soften the force by choosing a particle shape that is more spread out, dampening the effects of two-body interactions. A simple choice is that of a Plummer sphere~\citep{plummer1911problem};
\begin{align}
	S_{\text{P}}(\vec{x}) &\equiv \frac{3}{4\PI \epsilon^3}\biggl(1 + \frac{\vec{x}^2}{\epsilon^2}\biggr)^{-5/2} \label{eq:plummer_shape}  \\
	\Rightarrow |\vec{x}|_{\text{P}}^{-1} &= \bigl(\vec{x}^2 + \epsilon^2\bigr)^{-1/2}\,,
\end{align}
where $\epsilon \ge 0$ is the softening length, typically chosen to be a few percent of the mean inter-particle distance $L_{\text{box}}/\cbrt{N}$. Substituting $|\vec{x} - \vec{x}_{j\vec{n}}|^{-1}_{\pshape}$ for $|\vec{x} - \vec{x}_{j\vec{n}}|^{-1}_{\text{P}}$ into \eqref{eq:potential} and $|\vec{x}_{ij\vec{n}}|^{-3}_{\pshape}$ for $|\vec{x}_{ij\vec{n}}|^{-3}_{\text{P}} = (|\vec{x}_{ij\vec{n}}|^{-1}_{\text{P}})^3$ into \eqref{eq:force} then results in the Plummer softened potential and force, respectively.

Ideally, we would like the softening to vanish for large particle separation, i.e.\ we seek a shape with compact support. Though \CONCEPTONE{} implements both the point particle $S_\delta$ and the Plummer sphere $S_{\text{P}}$, the default softening shape is the B-spline of \citet{monaghan1985refined}, as also used by \GADGET{}:
\begin{align}
	S_{\text{B}}(\vec{x}) &\equiv 
	\frac{8}{\PI\epsilon_{\text{B}}^3} \begin{dcases}
			1 - 6x_{\text{B}}^2(1 - x_{\text{B}})  & \hspace{2.09em} x_{\text{B}} < \frac{1}{2}  \\
			2(1 - x_{\text{B}})^3                  & \frac{1}{2} \leq x_{\text{B}} < \makebox[\widthof{$\displaystyle \frac{1}{2}$}]{1}  \\
			0                        \hspace{7.6em} & \makebox[\widthof{$\displaystyle \frac{1}{2}$}]{1} \leq x_{\text{B}}
	\end{dcases}
	\label{eq:bspline_shape} \\
	\Rightarrow |\vec{x}|^{-1}_{\text{B}} &= \begin{dcases}
		\begin{aligned}
        \frac{32}{\epsilon_{\text{B}}}\biggl( &-\frac{1}{5}x_{\text{B}}^5 + \frac{3}{10}x_{\text{B}}^4 \\ &- \frac{1}{6}x_{\text{B}}^2 + \frac{7}{80} \biggr)\end{aligned} & \hspace{2.09em} x_{\text{B}} < \frac{1}{2} \\
        \begin{aligned}
        \frac{32}{\epsilon_{\text{B}}}\biggl( &\hspace{0.71em} \frac{1}{15}x^5_{\text{B}} - \frac{3}{10}x^4_{\text{B}} + \frac{1}{2}x^3_{\text{B}}  \\ &- \frac{1}{3}x^2_{\text{B}} + \frac{1}{10} - \frac{1}{480}x^{-1}_{\text{B}} \biggr)\end{aligned}
        & \frac{1}{2} \leq x_{\text{B}} < \makebox[\widthof{$\displaystyle \frac{1}{2}$}]{1}
        \\
        |\vec{x}|^{-1}  & \makebox[\widthof{$\displaystyle \frac{1}{2}$}]{1} \leq x_{\text{B}} \,,
	\end{dcases} \label{eq:bspline_potential} \\
	|\vec{x}|^{-3}_{\text{B}} &= \begin{dcases}
        \frac{32}{\epsilon_{\text{B}}^3}\biggl( \hspace{2.0em} x_{\text{B}}^3 - \frac{6}{5}x_{\text{B}}^2 + \frac{1}{3} \biggr)  & \hspace{2.09em} x_{\text{B}} < \frac{1}{2} \\
        \begin{aligned}
        \frac{32}{\epsilon_{\text{B}}^3}\biggl( &- \frac{1}{3}x_{\text{B}}^3 + \frac{6}{5}x_{\text{B}}^2 - \frac{3}{2}x_{\text{B}} \\ &+\frac{2}{3} - \frac{1}{480}x_{\text{B}}^{-3} \biggr)
        \end{aligned}  &\frac{1}{2} \leq x_{\text{B}} < \makebox[\widthof{$\displaystyle \frac{1}{2}$}]{1} \\
        |\vec{x}|^{-3}  \hspace{10.7em}  & \makebox[\widthof{$\displaystyle \frac{1}{2}$}]{1} \leq x_{\text{B}} \,, 
	\end{dcases} \label{eq:bspline_force}
\end{align}
where $x_{\text{B}} \equiv |\vec{x}|/\epsilon_{\text{B}}$ and $\epsilon_{\text{B}}$ is the B-spline softening length. Note that the symbol $|\vec{x}|^{-3}_{\text{B}} \neq (|\vec{x}|^{-1}_{\text{B}})^3$. Equations~\eqref{eq:bspline_potential,eq:bspline_force} then define the B-spline softened potential and force via \eqref{eq:potential,eq:force}, respectively. As in \citet{gadget2} we set $\epsilon_{\text{B}} = 2.8\epsilon$, keeping the Plummer softening length $\epsilon$ as the canonical softening parameter.

\paragraph*{Ewald summation}
The triply infinite sums of \eqref{eq:potential,eq:force} can be evaluated using the technique of~\citet{ewald1921berechnung} (see also~\citet*{hernquist1991application}). This amounts to writing the functional part of the potential \eqref{eq:potential} --- i.e.\ the reciprocal distance --- as a sum of a short-range and a long-range part; $|\vec{x}|^{-1} = {\mathcal{G}_{\text{sr}}}(\vec{x}) + {\mathcal{G}_{\text{lr}}}(\vec{x})$. We employ the common choice ${\mathcal{G}_{\text{sr}}}(\vec{x}) = \erfc(|\vec{x}|/[2x_{\text{s}}])|\vec{x}|^{-1}$, ${\mathcal{G}_{\text{lr}}}(\vec{x}) = \erf(|\vec{x}|/[2x_{\text{s}}])|\vec{x}|^{-1}$, where $x_{\text{s}}\ge 0$ is the short-/long-range force split scale. Transforming the long-range part to Fourier space\footnote{In an attempt to minimise notational clutter, Fourier-space quantities are distinguished from their real-space counterparts through their argument only.}, ${\mathcal{G}}_{\text{lr}}(\vec{k}) = 4\PI\exp(-x_{\text{s}}^2\vec{k}^2)/\vec{k}^2$, the potential may be written
\begin{equation}
	\varphi(\vec{x}) = -\frac{Gm}{a} \sum_{j=1}^N \left\{
	\begin{aligned}
	& \sum_{\vec{n}\in\mathbb{Z}^3} \left[
	\begin{aligned}
	    &\hspace{1.21em}\mathcal{G}_{\text{sr}}(\vec{x} - \vec{x}_{j\vec{n}}) \\
        &+ \bigl(|\vec{x} - \vec{x}_{j\vec{n}}|^{-1}_{\pshape} - |\vec{x} - \vec{x}_{j\vec{n}}|^{-1}\bigr)
    \end{aligned}
    \right]
    \\
	&\! - L_{\text{box}}^{-3} \sum_{\mathclap{ \vec{h} \in \mathbb{Z}^3\setminus \mathbf{0} }} \mathcal{G}_{\text{lr}} (\vec{k}_{\vec{h}}) \cos(\vec{k}_{\vec{h}} [\vec{x} - \vec{x}_j])
	\end{aligned}
	\right\}\,, \label{eq:ewald_potential}
\end{equation}
with $\vec{k}_{\vec{h}} \equiv 2\PI L_{\text{box}}^{-1}\vec{h}$. In \eqref{eq:ewald_potential} the softening is implemented by the parenthesis in the real-space sum over $\vec{n}$, ensuring that only the Newtonian part of the potential is softened, which decouples the choice of softening from the choice of how the potential has been split. For the B-spline softening \eqref{eq:bspline_potential} with compact support, this parenthesis vanishes for all images $\vec{n}$ of particle $j$ but that closest to $\vec{x}$, meaning that softening is only applied to the nearest image.

Figure~\ref{fig:geometry} depicts a simulation box with particles, along with various numerical aspects. For the top left particle, three single-particle potentials are shown: The unsoftened Newtonian potential $\propto |\vec{x}|^{-1}$, the softened Newtonian potential $\propto |\vec{x}|_{\text{B}}^{-1}$ and the softened short-range potential\footnote{The value of $x_{\text{s}}$ used for the short-range potential in Figure~\ref{fig:geometry} is one fitting for the \PTHREEM{} method (see section~\ref{subsubsec:p3m_gravity}), not for Ewald summation.} $\propto \mathcal{G}_{\text{sr}}(\vec{x}) + (|\vec{x}|_{\text{B}}^{-1} - |\vec{x}|^{-1})$. It is clearly seen how the softening removes the divergent behaviour in the vicinity of the particle --- without changing the potential further out for this case of B-spline softening --- and that the short-range potential tends to zero much more rapidly than the Newtonian potentials. We shall come back to Figure~\ref{fig:geometry} several times, referring to different aspects.

Given the Ewald prescription of the potential \eqref{eq:ewald_potential}, the comoving force on particle $i$ becomes
\begin{equation}
	\hspace{-0.8em}
	\vec{f}_i = -Gm^2 \sum_{\substack{j=1 \\ j \neq i}}^N \left\{
	\begin{aligned}
	& \sum_{\vec{n}\in\mathbb{Z}^3} \left[
		\begin{aligned}
		&\hspace{1.35em}|\vec{x}_{ij\vec{n}}|^{-3}
		\erfc\biggl(\frac{|\vec{x}_{ij\vec{n}}|}{2x_{\text{s}}}\biggr) \\
		&+ \frac{|\vec{x}_{ij\vec{n}}|^{-2}}{\sqrt{\PI}x_{\text{s}}}\exp\biggl(-\frac{\vec{x}_{ij\vec{n}}^2}{4x^2_{\text{s}}}\biggr) \\
    &+ \bigl(|{\vec{x}}_{ij\vec{n}}|^{-3}_{\pshape} - |\vec{x}_{ij\vec{n}}|^{-3}\bigr)
    \end{aligned}
    \right] \vec{x}_{ij\vec{n}}
    \\
	&\! + \frac{4\PI}{L_{\text{box}}^3}\,\, \sum_{\mathclap{\vec{h} \in \mathbb{Z}^3\setminus \mathbf{0} }}  \frac{\exp\bigl(-x_{\text{s}}^2\vec{k}_{\vec{h}}^2\bigr)}{\vec{k}_{\vec{h}}^2} \sin(\vec{k}_{\vec{h}} [\vec{x}_i - \vec{x}_j])\vec{k}_{\vec{h}}
	\end{aligned}
	\right\}\,, \label{eq:ewald_force}
\end{equation}
where again the softening term $(|{\vec{x}}_{ij\vec{n}}|^{-3}_{\pshape} - |\vec{x}_{ij\vec{n}}|^{-3})$ vanishes for all images $\vec{n}$ of particle $j$ but the one closest to particle $i$, in the case of B-spline softening.

The crux of the Ewald technique is that the infinite sums of \eqref{eq:ewald_potential,eq:ewald_force} converge exponentially, whereas the original infinite sums of \eqref{eq:potential,eq:force} converge much more slowly and in fact only conditionally \citep*{de1980simulation}. For some chosen $x_{\text{s}}$ the Ewald sums can then safely be truncated at some finite maximum $|\vec{n}|$ and $|\vec{h}|$. For the PP method \CONCEPT{} uses the values suggested by \citet{hernquist1991application};
\begin{equation}
\begin{dcases}
	x_{\text{s}} = \frac{L_{\text{box}}}{4} \,, \\
	|\vec{x}_{ij\vec{n}}| < 3.6L_{\text{box}} \,, \\
	\vec{h}^2 < 10\,,
\end{dcases} \label{eq:ewald_parameters}
\end{equation}
as do \GADGETTWO{}.

Despite having limited the infinite Ewald sums to a doable number of terms \eqref{eq:ewald_parameters}, the force computation for each particle pair $\{i, j\}$ --- corresponding to the large brace of \eqref{eq:ewald_force} --- is still substantial. In practice, \CONCEPT{} pre-computes this force for a cubic grid of particle separations between $0$ and $L_{\text{box}}/2$ in all three dimensions, with the softened contribution $|\vec{x}_{ij\vec{n}}|^{-3}_{\pshape}$ excluded. During simulation, forces are then obtained using CIC interpolation (covered in section~\ref{subsubsec:pm_gravity}) in this grid, with particle separations outside the tabulated region handled using symmetry conditions. The softened $|\vec{x}_{ij\vec{n}}|^{-3}_{\pshape} \vec{x}_{ij\vec{n}}$ from the nearest image is then added. By default, a grid size\footnote{Whenever the size of a (cubic) grid is given, it refers to the number of elements along each dimension. In case of the Ewald grid, this then consists of $64\times 64 \times 64$ elements (each containing a force vector).} of 64 is used for the Ewald grid.

\subsubsection{PM gravity}\label{subsubsec:pm_gravity}
Though the path towards the softened, Ewald-assisted periodic force~\eqref{eq:ewald_force} went through the potential $\varphi$, this potential itself is never actually computed by \CONCEPT{} when using the PP method. The particle-mesh (PM) method takes a different approach, establishing $\varphi$ as a cubic grid of size $n_{\varphi}$, from which particle forces are obtained via numerical differentiation and interpolation. The most expensive step of this method is the creation of $\varphi$, which in \CONCEPT{} is based on fast Fourier transforms (FFTs). Assuming (very reasonably) that the number of grid elements $n_{\varphi}^3\propto N$, the PM method then inherits the $\mathcal{O}(N \log N)$ complexity of the FFT~\citep{fft}, vastly outperforming the $\mathcal{O}(N^2)$ PP method. The price to pay is that of a limited resolution of gravity imposed by the finite size $L_{\varphi} = L_{\text{box}}/n_{\varphi}$ of the grid cells, which in practice is much larger than the particle softening length $\epsilon$ of the PP method.

We can explicitly solve the Poisson equation~\eqref{eq:Poisson} for the potential,
\begin{align}
	\varphi(\vec{x}) &= -Ga^2 |\vec{x}|^{-1} * \deltarho(\vec{x}) \\
	\Rightarrow \varphi(\vec{k}) &= -\frac{4\PI Ga^2}{\vec{k}^2}\deltarho(\vec{k}) \,, \label{eq:Poisson_Fourier}
\end{align}
where the convolution transforms to multiplication in Fourier space. The strategy of the PM method is to first interpolate the particle masses onto a grid, obtaining $\deltarho(\vec{x})$, then Fourier transforming this grid to obtain $\deltarho(\vec{k})$, converting to potential values $\varphi(\vec{k})$ through \eqref{eq:Poisson_Fourier}, then Fourier transforming back to real space, obtaining $\varphi(\vec{x})$. The same grid in memory is used to store all of these different quantities.

\paragraph*{Mesh interpolation}
As for the PP method, we wish to construct a density field $\rho(\vec{x})$ given the particle distribution by assigning a shape $S$ to the particles. Unlike direct summation, computing forces via the potential does not allow us to explicitly remove particle self-interactions, corresponding to the skipped $j=i$ terms of \eqref{eq:force,eq:ewald_force}. Instead, the particle shapes must be chosen such that any contributions from self-interactions vanish. For a cubic grid, this limits the possible shapes to the hierarchy \citep{HockneyEastwood}
\begin{equation}
	S_{p_{\text{i}}}(\vec{x}) = L_{\varphi}^{-3p_{\text{i}}} \bigast_{\mathclap{\text{$p_{\text{i}}$ times}}} \Pi\biggl(\frac{\vec{x}}{L_{\varphi}}\biggr) \,, \label{eq:shape_hierarchy}
\end{equation}
with the interpolation order $p_{\text{i}} \in \mathbb{N}_0$ and the big $\bigast$ operator representing repeated convolution. With the empty convolution understood to be the Dirac delta function, we obtain $S_0(\vec{x}) = \delta^3(\vec{x})$ as the lowest-order shape in the hierarchy. Higher-order shapes are then constructed through convolution with the cubic top-hat $\Pi(\vec{x}/L_{\varphi})$ spanning exactly one grid cell, with the top-hat function given by
\begin{equation}
	\Pi(\vec{x}) = \prod_{d=1}^3 \Pi\bigl(\vec{x}^{[d]}\bigr)\,, \qquad
	\Pi(x) = \begin{dcases}
		1 &  \hspace{2.09em}  |x| < \frac{1}{2} \\
		0 & 	\frac{1}{2} \le |x| \,,
	\end{dcases} \label{eq:tophat}
\end{equation}
where $\Pi$ of vector input is defined by multiplying results obtained from individual scalar inputs, $\vec{x}^{[d]}$ representing the $d$'th Cartesian scalar component of vector $\vec{x}$.

Given some interpolation order $p_{\text{i}}\ge 1$ we let the continuous density contrast field $\deltarho(\vec{x})$ be defined through \eqref{eq:density_contrast} with $S\rightarrow S_{p_{\text{i}} - 1}$, with $S_{p_{\text{i}} - 1}$ in turn given by \eqref{eq:shape_hierarchy}. We then define the discretised grid version of the density contrast $\deltarho_{\vec{m}}$ --- with $\vec{m}\in\mathbb{Z}^3$ labelling the mesh points at\footnote{Here vector-scalar addition is defined as adding the scalar to each element of the vector. Unlike e.g.\ \GADGET{}, \CONCEPTONE{} uses cell-centred grid values (by default), hence the offset by half a grid cell.} $\vec{x}_{\vec{m}} = L_{\varphi}(\vec{m} + \text{\textonehalf})$ --- via interpolation of the continuous $\deltarho(\vec{x})$ as follows:
\begin{align}
	\deltarho^{(1)}_{\vec{m}} &\equiv L_{\varphi}^{-3}\eval[3]{ \Pi\biggl(\frac{\vec{x}}{L_{\varphi}}\biggr) * \deltarho(\vec{x})}_{\vec{x} = \vec{x}_{\vec{m}}} \label{eq:density_grid_definition} \\
    &= \frac{m}{a^3} \sum_{\vec{n}\in\mathbb{Z}^3} \Biggl\{ -\frac{N}{L_{\text{box}}^{3}} + L_{\varphi}^{-3}\sum_{j=1}^N W_{p_{\text{i}}}\biggl( \frac{\vec{x}_{\vec{m}j\vec{n}}}{L_{\varphi}} \biggr) \Biggr\} \,, \label{eq:density_grid_implementation} \\
    &= \frac{m}{a^3} \sum_{\vec{n}\in\mathbb{Z}^3} \left\{
    \begin{aligned}
    &-\frac{N}{L_{\text{box}}^{3}} \\
    &+ \eval[4]{L_{\varphi}^{-3} W_{p_{\text{i}}}\biggl(\frac{\vec{x}}{L_{\varphi}}\biggr) * \sum_{j=1}^N \delta^3(\vec{x} - \vec{x}_{j\vec{n}})}_{\vec{x} = \vec{x}_{\vec{m}}}
    \end{aligned}
    \right\} \,, \label{eq:density_grid_deltaconvolution}
\end{align}
where we have introduced the dimensionless weight functions $W_{p_{\text{i}}}(\vec{x}/L_{\varphi}) \equiv L^3_{\varphi} S_{p_{\text{i}}}(\vec{x})$ and used $\vec{x}_{\vec{m}j\vec{n}} \equiv \vec{x}_{\vec{m}} - \vec{x}_{j\vec{n}} = \vec{x}_{\vec{m}} - (\vec{x}_j + L_{\text{box}}\vec{n})$. Equality~\eqref{eq:density_grid_implementation} is the one used for code implementation. The parenthesised superscript counts the number of particle $\leftrightarrow$ mesh interpolations carried out, which we shall want to keep track of.

\paragraph*{Deconvolved potential}
With the PM grid holding $\deltarho^{(1)}_{\vec{m}}$ values, an in-place FFT converts the values to $\deltarho^{(1)}_{\vec{h}}$, the grid version of $\deltarho(\vec{k})$ with $\vec{h}\in\mathbb{Z}^3$ labelling the grid points at $\vec{k}_{\vec{h}} = 2\PI L_{\text{box}}^{-1}\vec{h}$. This FFT treats the finite numerical representation of $\deltarho^{(1)}_{\vec{m}}$ as periodic, implementing the sum over images $\vec{n}$ of \eqref{eq:density_grid_implementation,eq:density_grid_deltaconvolution} automatically.

The density values are then converted to potential values using \eqref{eq:Poisson_Fourier}, resulting in grid values
\begin{equation}
	\varphi^{(1)}_{\vec{h}} = -\frac{4\PI Ga^2}{\vec{k}_{\vec{h}}^2} \deltarho^{(1)}_{\vec{h}}\,, \qquad \varphi^{(1)}_{\mathbf{0}} = 0\,,
\end{equation}
where the $\vec{k}=\mathbf{0}$ `DC' mode is explicitly zeroed, corresponding to removing the background density. This enables us to work with density values $\rho$ rather than density contrast values $\deltarho$ in the implementation, meaning we can ignore the subtraction of $N/L_{\text{box}}^3$ in \eqref{eq:density_grid_implementation,eq:density_grid_deltaconvolution}.

From \eqref{eq:density_grid_deltaconvolution} it is then clear that we can correct for the interpolation by dividing out the Fourier transformed weight function, allowing us to obtain deconvolved versions of the grid:
\begin{equation}
	\varphi^{(c)}_{\vec{h}} = \biggl[\frac{W_{p_{\text{i}}}(L_{\varphi}\vec{k}_{\vec{h}})}{L^3_{\varphi}}\biggr]^{c-1} \varphi^{(1)}_{\vec{h}} \,.
\end{equation}
The properly deconvolved potential grid is then given by $\varphi^{(0)}_{\vec{h}}$. Applying such deconvolution removes much of the spurious Fourier aliasing, improving the accuracy of the grid representation at small scales \citep{HockneyEastwood}.

\paragraph*{Obtaining forces}
We now transform back to real space using an in-place inverse FFT, obtaining $\varphi^{(c)}_{\vec{m}}$. We can then construct a force grid as
\begin{equation}
	\vec{f}^{(c)}_{\vec{m}} = -am L^{-1}_{\varphi} \vec{D}_{p_{\text{d}}}\varphi^{(c)}_{\vec{m}} \,,
\end{equation}
where $\vec{D}_{p_{\text{d}}}$ is some finite difference operator of order $p_{\text{d}}$. The resulting force grid $\vec{f}^{(c)}_{\vec{m}}$ must then be interpolated back to the particle positions and applied. Ignoring the sum over images $\vec{n}$ and subtraction of the background $N/L^3_{\text{box}}$ as previously mentioned, this interpolation is implemented by \eqref{eq:density_grid_implementation}, except that now the sum runs over mesh points instead of particle indices, as this time the interpolation is from the mesh onto the particles:
\begin{align}
	\vec{f}_i &= \sum_{\mathclap{\vec{m}\in\mathbb{Z}^3}} W_{p_{\text{i}}}\biggl(\frac{\vec{x}_i - \vec{x}_{\vec{m}}}{L_{\varphi}}\biggr) \vec{f}^{(-1)}_{\vec{m}} \label{eq:pm_force} \\
	&\begin{aligned}
		&= \frac{4\PI G m^2}{L^{4}_{\varphi}} \overbrace{
			\sum_{\mathclap{\phantom{\raisebox{1.9ex}{x}}}\mathclap{\phantom{\raisebox{-1.9ex}{x}}}\mathclap{
				\bigl\{ \, \vec{m} \, \big\thickvert \, |\vec{x}_i - \vec{x}_{\vec{m}}|_{\infty} < \frac{p_{\text{i}}L_{\varphi}}{2} \, \bigr\}
				}}
			W_{p_{\text{i}}}\biggl(\frac{\vec{x}_i - \vec{x}_{\vec{m}}}{L_{\varphi}}\biggr)
			}^{\smash{\text{particle $\leftarrow$ mesh}}}
			 \vec{D}_{p_{\text{d}}}
			 \,\underset{\smash{\hspace{-0.5em}\vec{m}\leftarrow\vec{h}}}{{\mathcal{F}_{\slashed{\mathbf{0}}}^{-1}}}
			 \Biggl\{ \\
		&\qquad \qquad
		\underbrace{\biggl[\frac{W_{p_{\text{i}}}(L_{\varphi}\vec{k}_{\vec{h}})}{L^3_{\varphi}}\biggr]^{-2}}_{\mathclap{\text{2 deconvolutions}}} 
		\smash{\overbrace{\frac{1}{\vec{k}^2_{\vec{h}}}}^{\mathclap{\text{Poisson kernel}}}}
		\hspace{0.3em}\underset{\smash{\mathclap{\vec{m}\rightarrow\vec{h}}}}{\mathcal{F}}\hspace{0.5em}
		\underbrace{\sum^N_{j=1} W_{p_{\text{i}}}\biggl(\frac{\vec{x}_{\vec{m}} - \vec{x}_j}{L_{\varphi}}\biggr)}_{\mathclap{\text{particles $\rightarrow$ mesh}}}
		 \Biggr\}\,,
	\end{aligned} \label{eq:pm_method}
\end{align}
where we specifically use $\vec{f}^{(-1)}_{\vec{m}}$ to take into account the additional particle $\leftarrow$ mesh interpolation, resulting in a total of 2 deconvolutions. The annotated equality~\eqref{eq:pm_method} provides a complete overview of the PM method by gathering up the different steps, with $\mathcal{F}$ representing the forward FFT and $\mathcal{F}_{\slashed{\mathbf{0}}}^{-1}$ the inverse FFT --- normalised so that $\mathcal{F}\mathcal{F}^{-1}\varphi_{\vec{m}} = \varphi_{\vec{m}}$ --- and the subscript indicating nullification of the $\vec{k}=\mathbf{0}$ mode prior to performing the inverse transform. The $\vec{m}\leftrightarrow \vec{h}$ below the FFT operators are just to indicate the change to the grid index caused by the transforms. Read this large expression backwards for it to follow the flow of the algorithm. For the localised weight functions $W_{p_{\text{i}}}$, the infinite sum over $\vec{m}$ in \eqref{eq:pm_force} only needs to be over mesh points in the vicinity of $\vec{x}_i$, as indicated for the sum over $\vec{m}$ in \eqref{eq:pm_method}, with $|\vec{x}|_{\infty} \equiv \max_d \bigl|\vec{x}^{[d]}\bigr|$ denoting the maximum norm. We shall look at $W_{p_{\text{i}}}$ in detail shortly, including how this particular definition of ``the vicinity'' arise.

In practice, the values stored in the PM grid goes through the transformations $\rho^{(1)}_{\vec{m}} \rightarrow \rho^{(1)}_{\vec{h}} \rightarrow \varphi^{(-1)}_{\vec{h}} \rightarrow \varphi^{(-1)}_{\vec{m}}$. A separate scalar grid is used to store the forces obtained from $\varphi^{(-1)}_{\vec{m}}$, along each dimension $d$ in turn. This scalar force grid is then interpolated onto all particles using \eqref{eq:pm_force}; $\{\vec{f}_i^{[d]}\} \leftarrow \vec{f}^{(-1)[d]}_{\vec{m}} = -am L^{-1}_{\varphi} \vec{D}^{[d]}_{p_{\text{d}}} \varphi^{(-1)}_{\vec{m}}$.

\paragraph*{Order of interpolation and differentiation}
Though the entire PM method is summarised by \eqref{eq:pm_method}, we have yet to explicitly write out the weight functions $W_{p_{\text{i}}}(\vec{x})$ and their Fourier transforms $W_{p_{\text{i}}}(\vec{k}_{\vec{h}})$ for different orders $p_{\text{i}}$. Similarly we have not yet specified the difference operators $\vec{D}_{p_{\text{d}}}$ for different orders $p_{\text{d}}$. We shall do so now.

From the definition $W_{p_{\text{i}}}(\vec{x}/L_{\varphi}) \equiv L_{\varphi}^3 S_{p_{\text{i}}}(\vec{x})$ along with \eqref{eq:shape_hierarchy}, the first weight functions are given by
\begin{align}
	W_{\text{NGP}}(x) &= \begin{dcases}
		1 	                      & \hspace{2.09em} |x| < \frac{1}{2} \\
		0 \hspace{7.565em}	     & \frac{1}{2} \le |x| \,,
	\end{dcases} \label{eq:W_NGP}  \\
	W_{\text{CIC}}(x) &= \begin{dcases}
		1 - |x|                   & \hspace{2.09em} |x| < \makebox[\widthof{$\displaystyle \frac{1}{2}$}]{1} \\
		0 \hspace{7.65em}         &                  \makebox[\widthof{$\displaystyle \frac{1}{2}$}]{1} \le |x| \,,
	\end{dcases} \label{eq:W_CIC}  \\
	W_{\text{TSC}}(x) &= \begin{dcases}
		\frac{3}{4} - x^2         & \hspace{2.09em}                |x| < \frac{1}{2} \\
		\frac{1}{8}(2|x| - 3)^2   &                \frac{1}{2} \le |x| < \frac{3}{2} \\
		0 \hspace{7.57em}         &                \frac{3}{2} \le |x| \,,
	\end{dcases} \label{eq:W_TSC}  \\
	W_{\text{PCS}}(x) &= \begin{dcases}
		\frac{1}{6}(3|x|^3 - 6x^2 + 4)	& \hspace{2.09em} |x| < \makebox[\widthof{$\displaystyle \frac{1}{2}$}]{1} \\
		\frac{1}{6}(2 - |x|)^3   		& \makebox[\widthof{$\displaystyle \frac{1}{2}$}]{1} \le |x| < \makebox[\widthof{$\displaystyle \frac{1}{2}$}]{2} \\
		0                         		& \makebox[\widthof{$\displaystyle \frac{1}{2}$}]{2} \le |x| \,,
	\end{dcases} \label{eq:W_PCS}
\end{align}
where common names --- `nearest grid point' (NGP) for $p_{\text{i}}=1$, `cloud in cell' (CIC) for $p_{\text{i}}=2$, `triangular shaped cloud' (TSC) for $p_{\text{i}}=3$, `piecewise cubic spline' (PCS) for $p_{\text{i}}=4$ --- have been used as labels. The behaviour regarding vector input is inherited from the top-hat~\eqref{eq:tophat}, i.e.\ $W_{p_{\text{i}}}(\vec{x})=W_{p_{\text{i}}}(\vec{x}^{[1]})W_{p_{\text{i}}}(\vec{x}^{[2]})W_{p_{\text{i}}}(\vec{x}^{[3]})$. All four weight functions are available in \CONCEPTONE{}. From \eqref{eq:W_NGP,eq:W_CIC,eq:W_TSC,eq:W_PCS} it is clear that grid points further away than $p_{\text{i}}/2$ grid units --- along any dimension --- from a particle's position do not take part in its interpolation; hence the set of grid points $\vec{m}$ included in \eqref{eq:pm_method}. In Figure~\ref{fig:geometry} the PM grid is drawn as thin grey lines, and the mass of the particle in the lower middle has been assigned to nearby grid points using PCS interpolation. The mass fractions are shown as assigned to the centres of the cells, reflecting the choice of cell-centred grid values in \CONCEPTONE{}.

As the hierarchy of real-space weight functions are generated through repeated convolution \eqref{eq:shape_hierarchy}, their Fourier transforms are generated through exponentiation (repeated multiplication). Given that $W_1(\vec{x}) = W_{\text{NGP}}(\vec{x})$ is just the top-hat~\eqref{eq:tophat}, we obtain
\begin{equation}
	\frac{W_{p_{\text{i}}}(L_{\varphi}\vec{k})}{L^3_{\varphi}} = \prod_{d=1}^3 \sinc^{p_{\text{i}}}\biggl(\frac{L_{\varphi}\vec{k}^{[d]}}{2}\biggr) \,, \label{eq:weight_fourier}
\end{equation}
with the cardinal sine function $\sinc(x) \equiv \sin(x)/x$ and we once more retain the same behaviour regarding vector input.

Now let us turn to the finite difference operator $\vec{D}_{p_{\text{d}}}$. This vector operator can be separated into three copies of the same scalar operator $\vec{D}_{p_{\text{d}}} = (D^{[1]}_{p_{\text{d}}}, D^{[2]}_{p_{\text{d}}}, D^{[3]}_{p_{\text{d}}})$, each acting along a separate dimension. The most natural choice is to use the optimally accurate symmetric difference approximation given the order $p_{\text{d}}$. If by $p_{\text{d}}$ we mean the number of grid points used for this approximation --- imposing $p_{\text{d}}\in 2\mathbb{N}$ due to the operation being symmetric --- these operators can be constructed as (see e.g.\ \citet{symmetric_diff})
\begin{equation}
	D_{p_{\text{d}}}\varphi_m = \sum_{\mathclap{\Delta m =-p_{\text{d}}/2}}^{p_{\text{d}}/2} \,\,\,\,\, \partial_{\xi} \,\,\,\,\, \prod_{\mathclap{\substack{\Delta m'=-p_{\text{d}}/2 \\ \Delta m' \neq \Delta m}}}^{p_{\text{d}}/2} \,\,\,\,\,\,  \eval[4]{\frac{\xi - \Delta m'}{\Delta m - \Delta m'}}_{\xi=0} \!\!\!\! \varphi_{m + \Delta m} \,, \label{eq:symmetric_diff_general}
\end{equation}
where the vector element superscript has been omitted and $\varphi_m$ is to be understood as a one-dimensional grid (or slice of the 3D grid $\varphi_{\vec{m}}$) with points labelled by $m\in\mathbb{Z}$ at $L_{\varphi}(m + \text{\textonehalf})$. \CONCEPTONE{} implements $p_{\text{d}}\in\{2, 4, 6, 8\}$, which from \eqref{eq:symmetric_diff_general} become
\begin{align}
	D_2\varphi_m &= \frac{1}{2} \biggl( \begin{aligned}
		&- \varphi_{m-1} \\
		&+ \varphi_{m+1}
	\end{aligned} \biggr) \,, \label{eq:symmetric_diff_2}  \\
	D_4\varphi_m &= \frac{1}{12} \biggl( \begin{aligned}
		&+ \varphi_{m-2} - 8\varphi_{m-1} \\
		&- \varphi_{m+2} + 8\varphi_{m+1}
	\end{aligned} \biggr) \,, \label{eq:symmetric_diff_4}  \\
	D_6\varphi_m &= \frac{1}{60} \biggl( \begin{aligned}
		&- \varphi_{m-3} + 9\varphi_{m-2} - 45\varphi_{m-1} \\
		&+ \varphi_{m+3} - 9\varphi_{m+2} + 45\varphi_{m+1}
	\end{aligned} \biggr) \,, \label{eq:symmetric_diff_6}  \\
	D_8\varphi_m &= \frac{1}{840} \left( \begin{aligned}
		&+ \phantom{16}3\varphi_{m-4} - \phantom{6}32\varphi_{m-3} \\
		&- \phantom{16}3\varphi_{m+4} + \phantom{6}32\varphi_{m+3} \\
		&+ 168\varphi_{m-2} - 672\varphi_{m-1} \\
		&- 168\varphi_{m+2} + 672\varphi_{m+1}
	\end{aligned} \right) \,, \label{eq:symmetric_diff_8}
\end{align}
with the symmetric property clearly manifest.

The interpolation order $p_{\text{i}}$ and difference order $p_{\text{d}}$ may be chosen independently, leading to many possible PM schemes available in \CONCEPTONE{}. By default, \CONCEPTONE{} uses $p_{\text{i}}=2$ (CIC) interpolation and $p_{\text{d}}=2$ differentiation for computing gravity via the PM method.

\paragraph*{Parallelisation}
We have yet to discuss the details of the MPI parallelisation of \CONCEPTONE{}, which necessarily must be integrated into the gravitational schemes. Given $n_{\text{p}}$ MPI processes, \CONCEPT{} divides the box into $n_{\text{p}}$ equally shaped cuboidal domains and assigns one such domain to each process. The exact domain decomposition chosen is uniquely\footnote{Up to permutation of the dimensions.} the one with the least elongated domains, minimizing the surface to volume ratio, in turn minimising communication efforts between processes. The domain decomposition shown in Figure~\ref{fig:geometry} --- with a thick black outline around each domain --- is for a simulation with $n_{\text{p}}=6$ processes, resulting in the decomposition $3\times 2\times 1$.

For the PP method, particles in one domain must explicitly be paired up with particles in all other domains. After having carried out the interactions of particles within their local domain, each process sends a copy of its particle data to another process --- the `receiver process' --- while simultaneously receiving particle data from a third process --- the `supplier process'. The interactions between local and received non-local particles are then carried out, with the momentum updates to the non-local particles sent back to the supplier process, while at the same time receiving and applying corresponding local momentum updates from the receiver process. This carries on for all such `dual' process/domain pairings, of which there are $\floor{n_{\text{p}}/2}$ from the point of view of any given local process, not counting the pairing between the local process and itself.

For the PM method the parallelisation efforts are more involved. The PM grid is distributed in real space according to the domains. Each grid cell must be entirely contained within a single domain, imposing the restriction that the number of domain subdivisions of the box along each dimension must divide $n_{\varphi}$. For the PM grid in Figure~\ref{fig:geometry}, $n_{\varphi}=54$ is chosen, which indeed is divisible by $3$, $2$ and $1$.

To carry out the required FFTs on the distributed grid, \CONCEPT{} employs the FFTW library \citep{fftw}, specifically its MPI-parallelised, real, 3D, in-place transformations. FFTW imposes a `slab'\footnote{Meaning distributed along a single dimension, resulting in local pieces of the global grid of shape $n_{\varphi}/n_{\text{p}}\times n_{\varphi}\times n_{\varphi}$.} decomposition of the global grid, in conflict with the cuboidal domain decomposition. Before performing a forward FFT, \CONCEPT{} then constructs a slab-decomposed copy of the domain-decomposed PM grid. Similarly, once the slab-decomposed grid is transformed back to real space, its values are copied over to the domain-decomposed grid. Furthermore, while in Fourier space, grids are transposed along the first two dimensions, as the last step in the distributed FFT routines is a global transposition, which is skipped for performance reasons. Similarly skipping this transposition step when transforming back to real space brings the dimensions back in order.

When it comes to particle interpolation using $W_{p_{\text{i}}}(\vec{x})$ \eqref{eq:W_NGP,eq:W_CIC,eq:W_TSC,eq:W_PCS} and grid differentiation using $\vec{D}_{p_{\text{d}}}$ \eqref{eq:symmetric_diff_2,eq:symmetric_diff_4,eq:symmetric_diff_6,eq:symmetric_diff_8}, data from a few (depending on the orders $p_{\text{i}}$ and $p_{\text{d}}$) grid cells away are required. Near a domain boundary, some of this required data belongs to a neighbouring domain and thus reside on a non-local process. To solve this, local domain grids are equipped with additional `ghost layers' of grid points surrounding the primary, local part of the grid. These ghost points must then be kept up-to-date with the corresponding non-local data, and vice versa. The required thickness $n_{\text{ghost}}$ of the ghost layers --- i.e.\ the number of ghost points extruding out perpendicular to a domain surface --- depends upon the orders $p_{\text{i}}$ and $p_{\text{d}}$. As already mentioned, \eqref{eq:W_NGP,eq:W_CIC,eq:W_TSC,eq:W_PCS} demonstrate that interpolation through $W_{p_{\text{i}}}(\vec{x})$ touches at most $p_{\text{i}}/2$ grid points to either side of a particle (along each dimension), thus requiring $n_{\text{ghost}} \ge \floor{p_{\text{i}}/2}$. For $\vec{D}_{p_{\text{d}}}$, the number of required ghost points can readily be read off of \eqref{eq:symmetric_diff_2,eq:symmetric_diff_4,eq:symmetric_diff_6,eq:symmetric_diff_8} as\footnote{Though rounding up $p_{\text{d}}/2$ is redundant for the symmetric difference operations of even order $p_{\text{d}}$, it becomes important for non-symmetric odd orders. \CONCEPT{} does in fact additionally implement $\vec{D}_1$, in both a `forward' and a `backward' version.} $n_{\text{ghost}} \ge \ceil{p_{\text{d}}/2}$. In total then,
\begin{equation}
	n_{\text{ghost}} = \max(\floor{p_{\text{i}}/2}, \ceil{p_{\text{d}}/2})
\end{equation}
ghost points are needed around local real-space domain grids.

Figure~\ref{fig:geometry} shows the ghost layers around the lower middle domain as ``ghostly'' shaded PM cells, using $n_{\text{ghost}} = 2$. As seen, the periodicity of the box is handled very naturally, which is really a secondary job almost automatically fulfilled by the ghost layers. Even in cases where the box is not subdivided along a given dimension, ghost layers are then still needed to implement the periodicity of the PM grid.

\subsubsection{\PTHREEMIT{} gravity}\label{subsubsec:p3m_gravity}
While the PM method is unrivalled in its performance, it comes with a severe limitation in resolution due to the finite grid cell size $L_{\varphi}$. One approach to overcome this is to only use PM for gravity at scales sufficiently large compared to $L_{\varphi}$, and then supply the missing short-range gravity using direct summation (PP) techniques. This hybrid PP-PM (\PTHREEM{}) method is the default gravitational solver of \CONCEPTONE{}. It comes with a free parameter $x_{\text{r}}$ which trades the accuracy of the PP method for the efficiency of the PM method, with practical values yielding a good balance.

\paragraph*{Combining PP and PM}
For the long-range part, the \PTHREEM{} method goes through all of the same steps as the PM method of section~\ref{subsubsec:pm_gravity}, with the Poisson kernel $4\PI/\vec{k}^2$ \eqref{eq:Poisson_Fourier} replaced with the long-range kernel $\mathcal{G}_{\text{lr}}(\vec{k}) = 4\PI \exp(-x_{\text{s}}^2\vec{k}^2)/\vec{k}^2$ \eqref{eq:ewald_potential} introduced earlier for the Ewald summation. Next, the missing short-range forces --- corresponding to the potential $\mathcal{G}_{\text{sr}}(\vec{x}) = |\vec{x}|^{-1}\erfc(|\vec{x}|/[2x_{\text{s}}])$ or the real-space sum over $\vec{n}$ of the force \eqref{eq:ewald_force} --- are added in using direct summation. Below, both the long-range and short-range sub-methods of the \PTHREEM{} method are spelled out:
\begin{equation}
		\begin{aligned}
		&
		\begin{drcases} \vec{f}_i =
		\frac{4\PI G m^2}{L^{4}_{\varphi}} \overbrace{
			\sum_{\mathclap{\phantom{\raisebox{1.9ex}{x}}}\mathclap{\phantom{\raisebox{-1.9ex}{x}}}\mathclap{
				\bigl\{ \, \vec{m} \, \big\thickvert \, |\vec{x}_i - \vec{x}_{\vec{m}}|_{\infty} < \frac{p_{\text{i}}L_{\varphi}}{2} \, \bigr\}
				}}
			W_{p_{\text{i}}}\biggl(\frac{\vec{x}_i - \vec{x}_{\vec{m}}}{L_{\varphi}}\biggr)
			}^{\smash{\text{particle $\leftarrow$ mesh}}}
			 \vec{D}_{p_{\text{d}}}
			 	\,\underset{\smash{\hspace{-0.5em}\vec{m}\leftarrow\vec{h}}}{{\mathcal{F}_{\slashed{\mathbf{0}}}^{-1}}}
			 \Biggl\{  \\
		\hspace{1.52em}
		\mathclap{\phantom{\raisebox{4.1ex}{x}}}
		\underbrace{\biggl[\frac{W_{p_{\text{i}}}(L_{\varphi}\vec{k}_{\vec{h}})}{L^3_{\varphi}}\biggr]^{-2}}_{\mathclap{\text{2 deconvolutions}}} 
		\smash{\overbrace{\frac{\exp\bigl(-x^2_{\text{s}}\vec{k}_{\vec{h}}^2\bigr)}{\vec{k}^2_{\vec{h}}}}^{\mathclap{\text{long-range kernel}}}}
			\underset{\smash{\mathclap{\vec{m}\rightarrow\vec{h}}}}{\mathcal{F}}\hspace{0.5em}
		\underbrace{\sum^N_{j=1} W_{p_{\text{i}}}\biggl(\frac{\vec{x}_{\vec{m}} - \vec{x}_j}{L_{\varphi}}\biggr)}_{\mathclap{\text{particles $\rightarrow$ mesh}}}
		 \Biggr\}
	\end{drcases} \rotatebox{-90}{\footnotesize{\hspace{-2.53em}long-range}}
	\\
	&\qquad \begin{drcases}
	\hspace{1.04em}-\hspace{0.4em}
	Gm^2\hspace{0.2em}
			\sum_{
				\mathclap{
					\substack{
						\{ \, j \, \thickvert \, |\vec{x}_{ij\vec{n}'}| < x_{\text{r}} \, \} \\
						j \neq i\mathclap{\phantom{\raisebox{0.8ex}{x}}}
					}
				}
			}
		\hspace{2.1em}
\left[
		\begin{aligned}
		&\hspace{1.35em}|\vec{x}_{ij\vec{n}'}|^{-3}
		\erfc\biggl(\frac{|\vec{x}_{ij\vec{n}'}|}{2x_{\text{s}}}\biggr) \\
		&+ \frac{|\vec{x}_{ij\vec{n}'}|^{-2}}{\sqrt{\PI}x_{\text{s}}}\exp\biggl(-\frac{\vec{x}_{ij\vec{n}'}^2}{4x^2_{\text{s}}}\biggr) \\
    &+ \phantom{\mathclap{|\vec{x}|^{-3}}}
    \smash{\underbrace{\bigl(|{\vec{x}}_{ij\vec{n}'}|^{-3}_{\pshape} - |\vec{x}_{ij\vec{n}'}|^{-3}\bigr)}_{\smash{\text{softening}}}}
    \end{aligned}
    \right] \vec{x}_{ij\vec{n}'}
	\end{drcases} \rotatebox{-90}{\footnotesize{\hspace{-2.53em}short-range}}
	\end{aligned}\label{eq:p3m_gravity}
\end{equation}
The exponential decay of the short-range force of \eqref{eq:p3m_gravity} allows us to only consider particle pairs within a distance $x_{\text{r}}$ a few times larger than $x_{\text{s}}$. In particular, choosing $x_{\text{s}}$ small compared to the box ensures that only the single image $\vec{n}'$ of particle $j$ nearest to particle $i$ has a non-negligible influence, ridding us of the sum over images $\vec{n}$. In \eqref{eq:p3m_gravity} then, $\vec{x}_{ij\vec{n}'} \equiv \vec{x}_i - \vec{x}_{j\vec{n}'} = \vec{x}_i - (\vec{x}_j + L_{\text{box}}\vec{n}')$ with $\vec{n}'$ chosen such that $|\vec{x}_{ij\vec{n}'}| = \min_{\vec{n}\in\mathbb{Z}^3} |\vec{x}_{ij\vec{n}}|$.

\begin{figure*}
\includegraphics[width=0.69\textwidth]{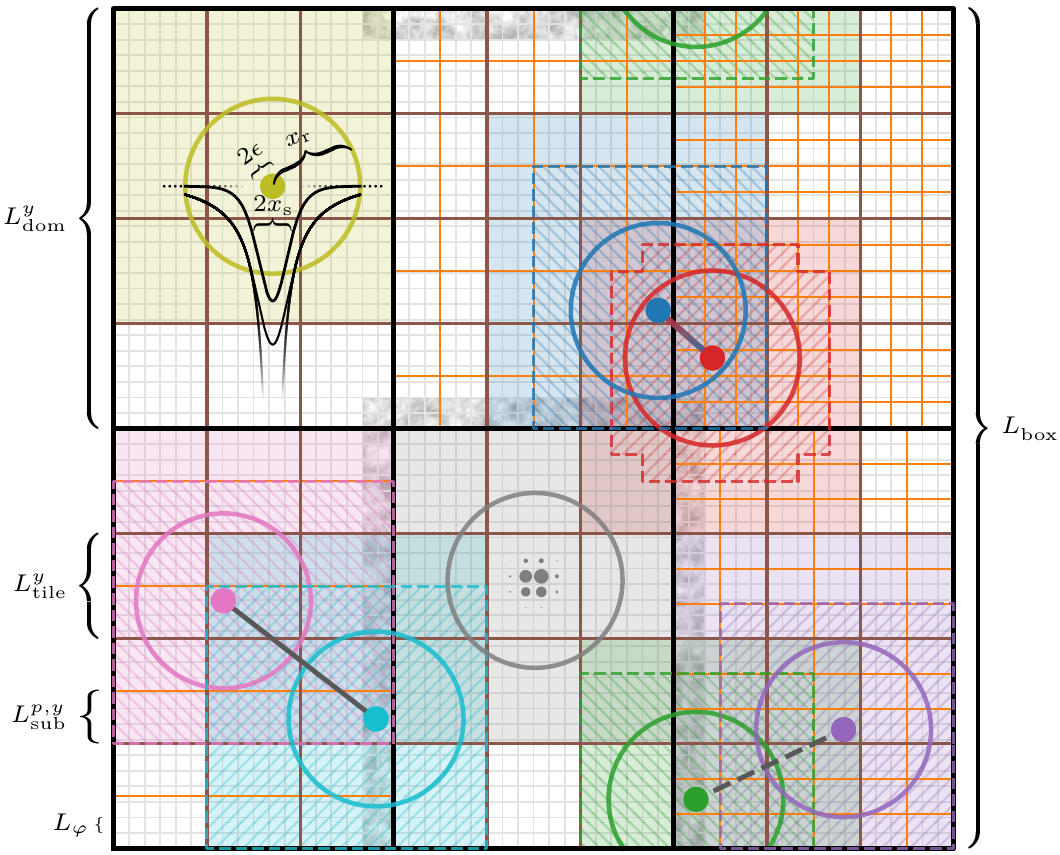}  
\caption{
The full geometric picture of the simulation box, demonstrating various aspects. In the example shown we imagine running a simulation with $n_{\text{p}} = 6$ processes, resulting in a domain decomposition of $3\times 2\times 1$. For clarity we shall ignore the last dimension. The entire cubic box (outer black square) of length $L_{\text{box}}$ is then subdivided into $3\times 2$ domains (black rectangles) all of size $L_{\text{dom}}^x\times L_{\text{dom}}^y$.
The global, cubic PM grid is shown in grey, with a grid size of $n_{\varphi} = 54$. A low number $N=8$ of particles is shown as small, different coloured solid circles with radii given by the softening length $\epsilon$ chosen as $0.03\, L_{\text{box}}/\cbrt{N}$. We note that the actual number of particles in a standard simulation of $n_{\varphi} = 54$ would be much greater.
The mass of the grey particle in the lower central domain is shown as being assigned to the PM grid cells using interpolation, specifically PCS \eqref{eq:W_PCS}. Also, ghost layers of the local PM grid using $n_{\text{ghost}} = 2$ are shown around this domain as ``ghostly'' shaded cells.
The raw Newtonian, softened Newtonian and softened short-range single-particle potentials are shown for the yellow particle at the upper left, with $x_{\text{s}}$ dictating the width of the localised short-range potential. As seen, the short-range potential is vanishingly small a distance $x_{\text{r}}$ away from the particle. The values \eqref{eq:p3m_parameters} are used for $x_{\text{s}}$ and $x_{\text{r}}$.
For two particles to interact under short-range gravity, they must be within $x_{\text{r}}$ of each other, i.e.\ each coloured hollow circle must contain (the centre of) the other particle. Thus here, only the blue and red particle pair interact.
The domains are subdivided into tiles of size $L^x_{\text{tile}}\times L^y_{\text{tile}}$, shown in brown. Each tile has to be at least $x_{\text{r}}$ along each dimension, here leading to a tile decomposition of $3\times 4$ of each domain. The $3\times 3$ tiles within reach of a given particle have been shaded with the colour of that particle, indicating neighbouring tiles needed to be checked for possible interacting partners of the given particle. Given this information alone, one can see that the upper right blue and red, the lower left magenta and cyan, as well as the lower right green and purple particles all have a chance of pairwise interacting.
Each tile is further subdivided into subtiles --- shown in orange --- independently within each domain. Their size are given by $L^{p,x}_{\text{sub}}\times L^{p,y}_{\text{sub}}$ with $p$ labelling the domain/process. Coloured hatched regions around particles show which subtiles are within reach $x_{\text{r}}$ of the subtile containing each particle. From this information, it is now clear that the green and purple particle are too far separated to interact via the short-range force. The subtile decomposition employed within the lower left domain is insufficient to tell us that the magenta and cyan particle do not interact.
}
\label{fig:geometry}
\end{figure*}

We seek to minimize $x_{\text{s}}$ in order to delegate as large of a fraction of the total work load as possible to the efficient PM part. Make $x_{\text{s}}$ too small however and the discrete nature of the grid will start to show up as spurious defects in the long-range force. The default\footnote{Whenever the term `default' is used in relation to \CONCEPT{}, we refer to default parameter values, all of which can be easily changed in parameter files, with no need for recompilation.} values employed by \CONCEPT{} for \PTHREEM{} is the same as those used by \GADGETTWO{} for TreePM:
\begin{equation}
\begin{cases}
	x_{\text{s}} = 1.25\, L_{\varphi} \,, \\
	x_{\text{r}} = 4.5\, x_{\text{s}} \,,
\end{cases} \label{eq:p3m_parameters}
\end{equation}
which is also what is depicted in Figure~\ref{fig:geometry}. Here $2x_{\text{s}}$ is shown for the upper left particle as dictating the width of the short-range potential, and a circle of radius $x_{\text{r}}$ is shown around every particle, illustrating their gravitational region of influence.

Using \eqref{eq:p3m_parameters}, the performance of the \PTHREEM{} method in \CONCEPTONE{} then depends on the grid size $n_\varphi$ through $L_{\varphi} = L_{\text{box}}/n_{\varphi}$. We prefer to run with
\begin{equation}
    n_{\varphi} = 2\cbrt{N}\,, \label{eq:phi_gridsize_preference}
\end{equation}
i.e.\ having 8 times as many PM cells as particles. While requiring quite a bit more memory than say $n_{\varphi} = 1\cbrt{N}$, this large cells to particles ratio lowers $x_{\text{s}}$, shifting a larger fraction of the computational burden onto the efficient long-range force, speeding up simulations significantly. Even so, for typical simulations the majority of the computation time is spent on the short-range forces, and so it is vital to implement these efficiently, to which we shall attend shortly.

As for the PM method of section~\ref{subsubsec:pm_gravity}, the \PTHREEM{} method in \CONCEPTONE{} employs $p_{\text{i}}=2$ (CIC) interpolation by default. As the long-range mesh of \PTHREEM{} is generally much smoother than the mesh of PM, it makes sense to increase the order of differentiation, and so $p_{\text{d}}=4$ is chosen as the default for \PTHREEM{} gravity in \CONCEPTONE{}. These default \PTHREEM{} settings of \CONCEPTONE{} thus coincide with the (fixed) TreePM settings of \GADGETTWO{}.

\paragraph*{Tiles}
What remains to be discussed is exactly how to efficiently implement the short-range\footnote{The perhaps equally complicated-looking \emph{long}-range sum over $\vec{m}$ of \eqref{eq:p3m_gravity} is in fact trivial to implement for our regular grid.} sum of \eqref{eq:p3m_gravity}, where each particle should be paired only with neighbouring particles within a distance $x_{\text{r}}$. What we need is to sort the particles in 3D space using some data structure, which then allows for efficient querying of nearby particles, given some location.

The data structure employed for the particle sorting in \CONCEPTONE{} is one we refer to as a \emph{tiling}. Here each domain is subdivided into as many equally sized cuboidal volumes --- called \emph{tiles}\footnote{Note that the word 'tile' is used by the \CUBEPTHREEM{} \citep{HarnoisDeraps:2012vd} and \CUBE{} \citep*{Yu_2018} codes as well, though to refer to a different kind of sub-unit, aiding with the parallelisation.} --- as possible, with the constraint that the tiles must have a size of at least $x_{\text{r}}$ along each dimension. This guarantees that a particle within a given tile only interacts with other particles in the surrounding $3\times 3\times 3$ block of tiles, i.e.\ with particles within its own tile or within the surrounding shell of 26 neighbouring tiles. The tiling is shown in brown on Figure~\ref{fig:geometry}, where the shape and size of the domains give rise to a tile decomposition $3\times 4\times 9$ (with the last dimension suppressed on the figure) of each domain. Note that since the domains are not cubic, the tiles will generally not be so either, as is the case on the figure. As all domains are equally shaped, all domain tilings will be similar, giving rise to a global (box) tiling. Further note that this global tiling generally do not align with the global PM mesh.

With the geometry of the tiling fixed, the particles are sorted into tiles in $\mathcal{O}(N)$ time. The interactions between particles within tiles are now carried out in a manner somewhat similar to the parallelisation strategy of the PP method described towards the end of section~\ref{subsubsec:pm_gravity}, though now both at the domain and at the tile level, below described separately for the two cases of tile interaction purely within the local domain and tile interaction across a domain boundary:
\begin{description}
	\item \textit{Local tile interaction}: Every process iterates over its tiles, in turn considering them as the `receiver tile'. After dealing with the interactions of particles within a given receiver tile itself, a neighbouring tile is selected as the `supplier tile'. Interactions between particles of the receiver and supplier tile are then carried out. A different neighbouring supplier tile within the local domain is then continually selected, until exhaustion. Once all neighbouring, local tiles (up to 26) have been dealt with, a different tile is considered as the receiver, and so on. Importantly, when selecting the next supplier tile, the one chosen must not have already been paired with the current receiver tile using opposite receiver/supplier roles.
	\item \textit{Non-local tile interaction}: The local process/domain is `dual-paired' with a non-local receiver and supplier process/domain, as in the PP method. Unlike the PP method, only the 26 neighbouring domains are considered, resulting in 13 pairings. The particles within local tiles neighbouring the receiver domain are sent to the receiver process, while corresponding particles are received from the supplier process. The local tiles neighbouring the supplier domain is then iterated over, in turn given the role as the receiver tile. Each such local receiver tile is then sequentially paired with non-local supplier tiles from the subset of the tiles (up to 9) received from the supplier process which neighbour the local receiver tile in question. Having directly updated the momenta of local particles due to the interactions, the non-local momentum updates are additionally sent back to the supplier process, while corresponding momentum updates are received from the receiver process, which are then applied as well.
\end{description}
Having at least 3 tiles across the box along each dimension ensures that the above scheme does not double count any tile pairs, even in extreme cases such as $n_{\text{p}} = 1$ where all 26 ``non-local neighbour domains'' are really all just the local domain itself. This constraint\footnote{In fact, \CONCEPTONE{} requires the global tiling to consist of at least 4 tiles across each dimension, as this simplifies some logic regarding the periodicity. For the standard values~\eqref{eq:p3m_parameters}, this restricts $n_{\varphi} \ge 23$ --- really $n_{\varphi} \ge 24$ as \CONCEPT{} further needs grid sizes to be even --- corresponding to $N\ge 12^3$ if we go with out default choice \eqref{eq:phi_gridsize_preference}, which is not much of a restriction at all.} is thus imposed by \CONCEPTONE{}.

With $n_{\text{tile}}$ the total number of tiles in the box, the average number of particles in a tile is $N/n_{\text{tile}}$, resulting in a time complexity for the tiled short-range force computation of $\mathcal{O}(N^2/n_{\text{tile}})$. As $n_{\text{tile}}\appropto n_{\varphi}^3$ this again shows how using a finer PM grid shifts the computational burden from the short-range computation over to the long-range computation. Furthermore, using $n_{\varphi}^3\propto N$, we see that the tiles formally reduce the full short-range interaction to linear time $\mathcal{O}(N)$, beating the rivalling $\mathcal{O}(N \log N)$ tree methods. In practice, having large inhomogeneities will make different tiles require different computational effort, degrading the performance. If the inhomogeneities extend to the domain scale, further degrading arises due to load imbalance between the processes, which \CONCEPT{} currently does not attempt to mend.

\paragraph*{Subtiles}
The basics of the tile-based short-range particle pairing has now been established, but it has room for optimisations. One such optimisation is that of \emph{sub}tiles, i.e.\ finer tiles within the main tiles.

In Figure~\ref{fig:geometry}, the circle of radius $x_{\text{r}}$ shown around every particle demonstrates the reach of the short-range force. In addition, the $3\times 3$ block of tiles surrounding a particle is shaded with a colour matching that particle, showing the possible tiles in which interacting partner particles might reside. Though in fact only the blue and red particles in the upper right are close enough to interact, the magenta and cyan pair in the lower left as well as the green and purple pair in the lower right seems like equally good candidates for possible interaction, from the point of view of the tiling.

Once two particles $i$ and $j$ have finally been paired up by the tiling mechanism, their separation $|\vec{x}_i - \vec{x}_j|$ is measured, upon which the interaction is aborted if $|\vec{x}_i - \vec{x}_j| > x_{\text{r}}$. For ideal small, cubic tiles of volume $x^3_{\text{r}}$, this happens for $\SI{9}{\percent}$ of interactions with both particles within the same tile, for $\SI{66}{\percent}$ of interactions between tiles sharing a face, for $\SI{91}{\percent}$ of interactions between tiles sharing only an edge, and for $\SI{98}{\percent}$ of interactions between tiles sharing only a corner. The reason for adding subtiles is to exclude many of these non-interactions early, accelerating the short-range computation. This is done by extending the tile pairing mechanism with subtiles, which are likewise paired up. Crucially, only subtiles so near each other that they could potentially contain interacting particles are paired, leading each receiver subtile to be paired up with subtiles within a surrounding, blocky ball, approaching a smooth ball of radius $x_{\text{r}}$ in the limit of infinite subtiles.

Unlike the main tiles, subtiles are local to each domain, meaning that each process is free to choose its own subtile decomposition, though with the same employed throughout its domain. Figure~\ref{fig:geometry} shows a variety of subtile decompositions in orange, e.g.\ $2\times 3$ for the lower right domain. Here we also find the green and purple particle, which according to the tiling needs to be paired, as the green particle is within the purple shaded region, and vice versa. The green and purple hatching shows which subtiles are reachable from the particular subtile containing each particle. As the hatched regions do not contain the subtile of the other particle, it means that adding in this subtiling indeed saves us from having to consider this irrelevant particle pair. Turning to the lower left domain of Figure~\ref{fig:geometry}, we see that the applied subtile decomposition of $1\times 2$ is insufficient to rid us of the irrelevant pairing of the magenta and cyan particle, even though their separation is more than twice the critical distance $x_{\text{r}}$. Increasing the number of subdivisions by just 1 in either dimension would have made the difference.

Finally let us consider the blue and red particle pair at the upper right of Figure~\ref{fig:geometry}, where no amount of subtiling will reject the pairing since these particles are close enough for interaction to take place. For the domain containing the red particle, a subtile decomposition of $3\times 4$ is used, which is substantial enough for the red hatched region to become slightly blocky. As the two particles reside on different processes, the interaction cannot take place before one of the processes sends its particle to the other process, as described earlier. The received particle(s) are then re-sorted according to the local subtiling, which is then traversed in order to locate particle pairs. This means that the subtile decomposition used for the $\text{blue}\leftrightarrow\text{red}$ interaction depends upon which process ends up as being considered the receiver and which the supplier. This is why Figure~\ref{fig:geometry} shows the blue and red hatched regions extending into the other domain, disregarding the different subtiling used here. Though the details of the inter-process communication may then affect the number of paired particles, which particle pairs end up interacting in the end remain unaffected.

Though subdividing space further could always lead to a still lower number of mistakenly paired particles, the overhead associated with the increased number of subtiles means that a sweet spot exists. Generally, higher particle number densities call for finer subtile decompositions. By default, \CONCEPTONE{} automatically estimates the optimal subtiling within each domain. Over time, each process periodically checks whether it is worth subdividing further due to the increased inhomogeneity. It does so by temporarily applying a slightly more refined subtile decomposition and comparing the measured time for a short-range force computation with a record of previous such computation times. If superior, the refined subtiling is kept, otherwise the old one is switched back in. The subtiles are thus both spatially and temporally adaptive.

\paragraph*{Other optimisations}
\CONCEPTONE{} goes to great lengths in order to arrive at a performant short-range computation, as evident from the implementation of subtiles, including automatic refinement. Here we briefly want to discuss further such short-range optimisations employed.

The two-level tile \textplus{} subtile structure is reminiscent of a shallow tree. While the geometry of a full tree reflects the underlying particle structure, the geometry of our (sub)tilings is determined solely by the simple Cartesian subdivisions. This allows us to pre-compute which of the (sub)tiles to pair with each other, eliminating a lot of decision making from within the actual `walk' (the iteration over $\text{tiles}\rightarrow\text{subtiles}\rightarrow\text{particles}$), which in turn saves on clock cycles and lowers the pressure on the branch predictor. Having a static, non-hierarchical data structure further results in simple access patterns with minimal pointer chasing, allowing for proper exploitation of CPU cache prefetching.

Once two particles $i$ and $j$ have been selected for interaction, the first thing to do is to compute their mutual squared\footnote{We keep working with \emph{squared} distances in order not to perform an expensive square root operation.} distance $|\vec{x}_{ij\vec{n}'}|^2$, after which the interaction is rejected if $|\vec{x}_{ij\vec{n}'}|^2 \ge x^2_{\text{r}}$, in accordance with the short-range sum of \eqref{eq:p3m_gravity}. Here we need to effectively shift $\vec{x}_i - \vec{x}_j$ by $L_{\text{box}}\vec{n}'$ as to minimise $|\vec{x}_{ij\vec{n}'}|^2 = |\vec{x}_i - \vec{x}_j - L_{\text{box}}\vec{n}'|^2$, corresponding to finding the image of particles $j$ nearest to particle $i$. The solution $\vec{n}'$ can in fact be determined just from knowing the tiles of particle $i$ and $j$, and so we pre-compute this already at the tile pairing level. In the typical case of many tiles across the box, the vast majority of tile pairs will have $\vec{n}' = \mathbf{0}$. To take advantage of this, explicit loop unswitching\footnote{This is achieved through custom transpiler directives and code transformations, briefly introduced in section~\ref{subsec:language}.} is utilised to completely eliminate the redundant zero-shift in these cases.

With particles $i$ and $j$ finally selected and deemed close enough for interaction to occur, we now need to compute their mutual short-range force, given by the large bracket of \eqref{eq:p3m_gravity}. Given that it is needed within the tightest loop of the program, this large expression is quite expensive. We thus have it (including the softening term) tabulated in a 1D table, indexed by $|\vec{x}_{ij\vec{n}'}|^2$ between $0$ and $x^2_{\text{r}}$. Here we use the cheapest possible (1D) NGP lookup, with the table being rather large\footnote{By default, this table has $2^{12}$ elements. A far smaller table and e.g.\ linear interpolation would work as well, but at the cost of performance.} in order to ensure accurate results nonetheless. This strategy works well for modern hardware with large CPU caches.

To further enable good utilisation of the CPU caches, the particles are ordered in memory in accordance with the visiting order resulting from the $\text{tile}\rightarrow\text{subtile}\rightarrow\text{particle}$ walk. The spatial drifting of the particles will gradually degrade this previously optimal sorting, and so the in-memory reordering of the particles is periodically reapplied.

\paragraph*{Recap of subvolumes}
The simulations of \CONCEPTONE{} make use of several different, nested subvolumes, in particular when using \PTHREEM{}. It may not be clear why we need this many levels of nested subvolumes, or indeed why we do not opt for even more. In fact, each such level exists for a very particular reason, which is briefly outlined below.

\begin{description}
	\item \textit{Box}: Though usually not thought of as a \emph{sub}volume, the simulation box itself exists in order to reduce an infinite universe to a finite volume with a finite number of degrees of freedom. The infinity of space is then imitated by the imposed periodicity.
	\item \textit{Domains}: The box is subdivided into domains in order to reduce the $N$-body problem into parallelisable chunks, to be distributed over many CPUs. A one-to-one mapping between domains, CPU cores and MPI processes is used within \CONCEPT{}.
	\item \textit{Tiles}: The domains are subdivided into tiles in order to take advantage of the finite range of the short-range force, partitioning the particles into subvolumes with the guaranteed property that particles within one such subvolume does not interact with particles further away than the nearest neighbour subvolumes. In particular, this lends itself to easy, near-minimal communication between processes.
	\item \textit{Subtiles}: Subtiles exist purely as an optimisation layer, accelerating the short-range computation through effective early rejection of particle pairs, by corresponding elimination of subtile pairs. Unlike all other subvolumes, the numbers of subtiles are free to change over time, adapting to increased inhomogeneity.	In addition, since subtiles are never shared between processes, the number of subtiles is free to vary from domain to domain, introducing spatial adaptivity as well.
\end{description}

One can imagine introducing a still deeper level of subvolumes, i.e.\ `subsubtiles', with the hope of further speeding up the computation. For this to not be equivalent to simply increase the number of subtiles, the coarseness of the subsubtilings would have to vary across the domain, e.g.\ within each tile or subtile. This would in turn imply that the subvolume geometry considered by a given process varies from place to place, which will decrease CPU cache performance. On top, there of course comes a point where sorting particles into still finer subvolumes and indexing into them outweigh the benefits from slightly increased early particle pair rejection. Given a large enough number of processes $n_{\text{p}}$, the spatial adaptiveness of the subtilings ensures that this in fact is the optimal level at which to stop subdividing space. We conjecture that this is the case also for typical values of $n_{\text{p}}$.

\subsection{Time-stepping}\label{subsec:timestepping}
This subsection describes the time-stepping mechanism implemented in \CONCEPTONE{}, including how the global simulation time step is chosen throughout cosmic history, and how this global time step is subdivided into finer steps, generating adaptive particle time-stepping.

As alluded to in section~\ref{subsec:gravity}, \CONCEPT{} employs cosmic time $t$ as its choice of time integration variable, and makes use of comoving coordinates $\vec{x}\equiv \vec{r}/a$ --- with $\vec{r}$ being physical coordinates --- and associated canonical momenta $\vec{q}\equiv a^2m\dot{\vec{x}}$ with $\dot{\phantom{x}}\equiv\partial_t$. The Hamiltonian equations of motion for the particles are then \citep{Peebles1980}
\begin{equation}
	\begin{dcases}
		\dot{\vec{x}}_i(t) = \frac{\vec{q}_i(t)}{a^2(t)m} \,, \\
		\dot{\vec{q}}_i(t) = \frac{\vec{f}_i(t)}{a(t)} \,,
	\end{dcases} \label{eq:eom}
\end{equation}
with the comoving force $\vec{f}_i$ being the primary subject of section~\ref{subsec:gravity}.

Given the state of the $N$-body system $(\{\vec{x}_i(t)\}, \{\vec{q}_i(t)\})$ at some time $t$, the coupled\footnote{Remember that $\vec{f}_i$ depends explicitly on all positions $\{\vec{x}_{j\neq i}\}$.} equations \eqref{eq:eom} can be solved numerically by alternatingly evolving $\{\vec{x}_i(t)\} \rightarrow \{\vec{x}_i(t + \Delta t)\}$ (keeping $\{\vec{q}_i\}$ fixed) and $\{\vec{q}_i(t)\} \rightarrow \{\vec{q}_i(t + \Delta t)\}$ (keeping $\{\vec{x}_i\}$ fixed) over discrete time steps of size $\Delta t$.

\subsubsection{Global time step size}\label{subsubsec:global_time_step_size}
\begin{figure}
\includegraphics[width=\columnwidth]{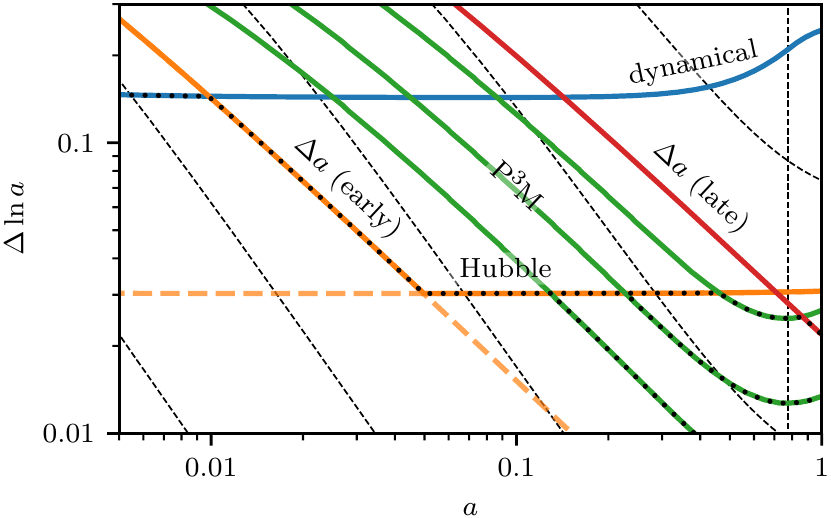}
\caption{Primary time step limiters in \CONCEPTONE{} and resulting time step size, expressed as $\Delta \ln a = \Delta a(t)/a(t) = \ln a(t + \Delta t) - \ln a(t)$, as function of scale factor $a$. Different limiters dominate at different times, as indicated by the dotted path, showing the evolution of the time step size itself. Note that `$\Delta a$ (early)' and `Hubble' are part of the same limiter, with the limiter value chosen as the maximum of the two sub-limiter values. Most limiters depend solely on the background cosmology, the exception being the \PTHREEM{} limiter which depends on the particle dynamics and thus the simulation resolution. The \PTHREEM{} limiter is shown for the cases $L_{\text{box}} \in \{2, 1, \text{\textonehalf}\}\SI[parse-numbers = false]{\cbrt{N}}{Mpc}/h$, with smaller box sizes (higher resolution) leading to lower allowed time step sizes. The dotted path is shown going through all three cases, though of course only a single \PTHREEM{} limiter exists for a given simulation. The qualitative change in behaviour of some of the limiters at late times is caused by the transition to $\Lambda$ domination, with matter-$\Lambda$ equality indicated by the black vertical dashed line. A standard $\Lambda$CDM cosmology (see Table~\ref{tab:cosmo_params}) was used to produce the figure, with all not-too-exotic cosmologies resulting in similar limiter values. Though $\Delta \ln a$ decreases over time, the time step size $\Delta t$ generally increases since $\Delta t\propto a^{3/2}\Delta \ln a$ in matter domination. The black slanted dashed lines represent trajectories of constant $\Delta t$, which indeed generally have steeper slopes than the coloured limiter lines.
}
\label{fig:timestepsize}
\end{figure}

Typical cosmological $N$-body simulations start from initial conditions at early, linear times (say $t\approx\SI{10}{Myr}$) and evolve the system forward to the present, non-linear time (say $t\approx\SI{14}{Gyr}$). During this evolution, physical phenomena --- related to the particles themselves as well as the cosmological background --- and numerical aspects introduce various time scales, above which the discrete time-stepping cannot operate if we are to hope for a converged solution. This leads to the concept of a time step limiter; a condition imposing a maximum allowed value for $\Delta t$, given by a small fraction of a corresponding time scale. Below we list the main such limiters (time scales) implemented in \CONCEPTONE{}, shown together in Figure~\ref{fig:timestepsize}.

\begin{description}
	\item \textit{Dynamical}: The gravitational dynamical time scale $(G\bar{\rho})^{-1/2}$, with $\bar{\rho}$ the background density of all non-linear components in the simulation.
	\item \textit{Fixed $\Delta a$ (late)}: The time step $\Delta t$ corresponding to a fixed $\Delta a$.
	\item \textit{Fixed $\Delta a$ (early) and the Hubble time}: This limiter is constructed as the maximum of two sub-limiters; the value $\Delta t$ which corresponds to a fixed $\Delta a$, and the instantaneous Hubble time $H^{-1}(t)$.
	\item \textit{\PTHREEMIT{}}: The time it takes to traverse a distance equalling the short-/long-range force split scale $x_{\text{s}}$ given the root mean square velocity of the particle distribution, $x_{\text{s}}/\sqrt{\langle \dot{\vec{x}}^2 \rangle}$.
\end{description}

As seen from Figure~\ref{fig:timestepsize}, which limiter dominates is subject to change during typical simulations. Of the above, only the \PTHREEM{} limiter is non-linear --- meaning it depends on the particle system --- with higher particle resolutions leading to a smaller maximal allowed $\Delta t$. All other limiters listed are obtained solely from the background.

\begin{figure*}
\includegraphics[width=0.88\textwidth]{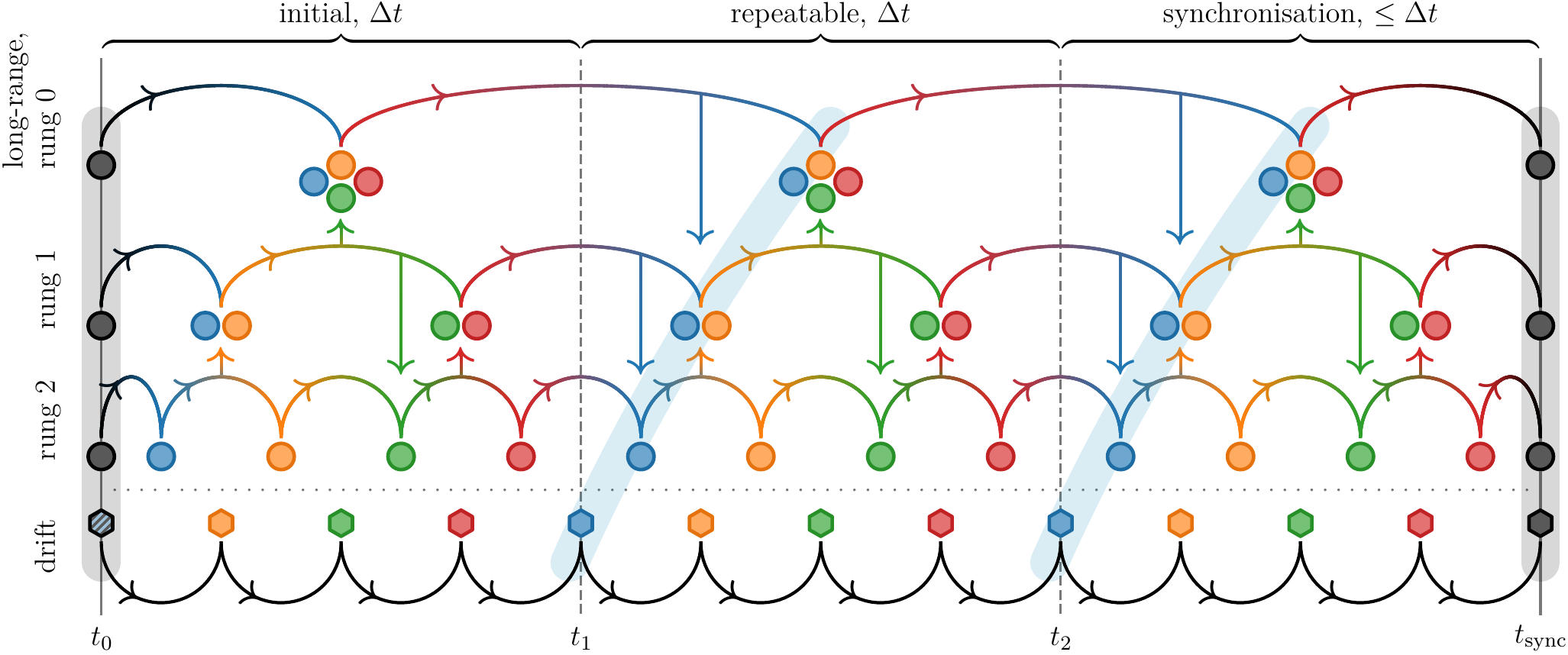}  
\caption{
Time-stepping scheme of \CONCEPTONE{} using rung-based leapfrogging. The series of connected hexagons indicates the discrete timeline followed by the particle positions $\{\vec{x}_i\}$ as they are evolved through drift operations. Note that all positions remain mutually synchronised throughout time. The momenta $\{\vec{q}_i\}$ are distributed among the different rungs, each with its own discrete timeline indicated by horizontally connected circles. For clarity, we consider the case of only $n_{\text{rung}} = 3$ rungs.
At the initial time $t_0$, all rungs are synchronised mutually and with the positions, as indicated by the vertical band covering the dark circles and the dark (and blue) hexagon. An initial `half' kick of size $\Delta t/2^{\ell+1}$ is now applied within each rung $\ell$, evolving them forward in time to the blue circles. As kicks of different rungs commute (Newtonian gravitation has no dependency on momentum), the order in which these kicks are applied does not matter. Considering the two-coloured hexagon now as blue, the whole system is now in a blue state.
Through rung-based kicks and collective drifts, the system now changes state from $\text{blue}\rightarrow\text{yellow}\rightarrow\text{green}\rightarrow\text{red}\rightarrow\text{blue}$, each cycle amounting to a full global time step $\Delta t$. We have $2^{n_{\text{rung}}-1}=4$ types (colours) of states due to the choice of $n_{\text{rung}} = 3$.
The changes of state respect the leapfrogging scheme for all rungs and are self-similar from one rung to the next: From blue to yellow, one drift of size $\Delta t/2^{\ell \geq n_{\text{rung}} - 1}$ is followed by a similar sized kick, only applicable to the highest rung $\ell = n_{\text{rung}} - 1 = 2$. All lower rungs evolve trivially from blue to yellow. $\text{Yellow}\rightarrow \text{green}$ consists of a similar drift followed by a kick of size $\Delta t/2^{\ell \geq n_{\text{rung}} - 2}$ in the highest two rungs $\ell\in\{2, 1\}$. $\text{Green}\rightarrow\text{red}$ is similar to $\text{blue}\rightarrow\text{yellow}$. $\text{Red}\rightarrow\text{blue}$ consists of the usual drift followed by a kick of size $\Delta t/2^{\ell \geq n_{\text{rung}} - 3}$, i.e.\ a kick to all rungs $\ell \in \{2, 1, 0\}$.
Once back at a blue state, a full time step $\Delta t$ has been completed, as indicated by the tilted blue band. Note though that only the positions are truly at $t_1 = t_0 + \Delta t$, while the momenta are all ``half a step ahead''. Synchronisation at some arbitrary time $t_{\text{sync}}$ is achieved simply by restricting the drifts and kicks to not evolve past this time, while otherwise keeping the scheme as is.
Once synchronised, all rungs are recomputed and assigned.
After a kick within a rung, some particles may accelerate enough so that they no longer belong within their given rung, in accordance with \eqref{eq:rung}. Such particles jump to a more appropriate neighbouring rung by making their next kick either $\text{\textonehalf}$ or $\text{\textthreequarters}$ as large as usual, as indicated by the vertical arrows. Jumping to a lower rung is only possible at every other kick.
In the above, `kicks' really refer to momentum updates due to short-range forces only, i.e.\ the lower half of \eqref{eq:p3m_gravity}. The long-range forces are applied to all particles whenever rung 0 is kicked.
}
\label{fig:timestepping}
\end{figure*}

Studying the linear growth of matter perturbations in a matter-dominated universe, we have $D\propto a$, $D=D(a)$ being the growth factor \citep{heath77,Peebles1980}. A fixed relative tolerance on the discrete evolution of $D$ is then ensured if we keep $\Delta D/D \propto \Delta a/a = \Delta \ln a$ constant. As evident from Figure~\ref{fig:timestepsize} this is equivalent to having $\Delta t \propto H^{-1}$, i.e.\ the Hubble limiter. This limiter is employed by \GADGETTWO{} all the way from early times until non-linear limiters take over. We have found that this leads to unnecessarily fine time steps early on, probably due to the very simple initial conditions with each particle coasting along a nearly straight path. While \GADGETTWO{} includes the horizontal dashed line of Figure~\ref{fig:timestepsize} as part of its Hubble limiter, \CONCEPTONE{} effectively changes the `fixed' value of $\Delta \ln a$ by instead using the dynamical limiter at early times, employing the constant $\Delta a$ (early) as a bridge between the two.

\CONCEPTONE{} implements a few extra limiters, which only come into play for non-standard simulations. These include a non-linear PM limiter and a non-linear Courant limiter for fluid components, as well as component-wise background limiters for the relativistic transition time for components with changing equation of state (relevant for e.g.\ non-linear massive neutrinos, see~\citet{nuconcept}) and for the life time of decaying components (relevant for decaying matter, see~\citet{concept_dcdm}).

For minimal loss of symplecticity during time-stepping (described in section~\ref{subsubsec:adaptive_timestepping}), the time step size $\Delta t$ should be kept constant over many steps. On the other hand, keeping $\Delta t$ at a lower value than necessary introduces further steps than required given the target accuracy. In \CONCEPT{} we use a period of 8 steps\footnote{Beyond striking a good balance, a period of 8 steps plays well with the non-linear fluid implementation as described in \citet{nuconcept}. Should the maximum allowed value of $\Delta t$ \emph{de}crease below its current value, the current period is terminated early.}, after which the particle system is synchronised (see section~\ref{subsubsec:adaptive_timestepping}) and $\Delta t$ allowed to increase in accordance with the limiters.

\subsubsection{Adaptive particle time-stepping}\label{subsubsec:adaptive_timestepping}
With the size of the time step $\Delta t$ determined, \CONCEPT{} integrates the particle system forwards in time using a symplectic second-order accurate leapfrog scheme~\citep{quinn1997time}, as is typical for $N$-body simulations. This is implemented using drift and kick operators $D$ and $K$, which advance the canonical variables as $\{\vec{x}_i(t)\} \xrightarrow{D(\Delta t)} \{\vec{x}_i(t+\Delta t)\}$, $\{\vec{q}_i(t)\} \xrightarrow{K(\Delta t)} \{\vec{q}_i(t+\Delta t)\}$. Discretising \eqref{eq:eom}, their implementations become\footnote{Importantly, $\vec{f}_i$ itself has no explicit dependence on $a$, as seen from e.g.\ \eqref{eq:p3m_gravity}.}
\begin{align}
    \begin{pmatrix*}[l]\displaystyle
        \{\vec{x}_i(t)\} \\ \displaystyle
        \{\vec{q}_i(t)\}
    \end{pmatrix*}
    &\xrightarrow{D(\Delta t)}
    \begin{pmatrix*}[l]\displaystyle
        \biggl\{\vec{x}_i(t) + \frac{\vec{q}_i(t)}{m}\int_t^{t + \Delta t}\!\frac{\d t'}{a^2(t')} \biggr\} \\
        \{\vec{q}_i(t)\}
    \end{pmatrix*}\,,
    \\
    \begin{pmatrix*}[l]\displaystyle
        \{\vec{x}_i(t)\} \\ \displaystyle
        \{\vec{q}_i(t)\}
    \end{pmatrix*}
    &\xrightarrow{K(\Delta t)}
    \begin{pmatrix*}[l]\displaystyle
        \{\vec{x}_i(t) \} \\ \displaystyle
        \biggl\{\vec{q}_i(t) + \vec{f}_i(t)\int_t^{t + \Delta t}\!\frac{\d t'}{a(t')} \biggr\}
    \end{pmatrix*} \,.
\end{align}
To evolve the synchronised system $(\{\vec{x}_i(t)\}, \{\vec{q}_i(t)\})$ it is first desynchronised by applying $K(\Delta t/2)$. The system is then evolved through repeated application of $D(\Delta t)$ followed by $K(\Delta t)$, under which $\{\vec{x}_i\}$ and $\{\vec{q}_i\}$ take turns leapfrogging past each other in time. Re-synchronisation of the canonical variables is achieved by some final drift and kick of appropriate size less than or equal to $\Delta t$.

\paragraph*{Individual time steps}
As the non-linear \PTHREEM{} time step limiter of Figure~\ref{fig:timestepsize} is set through the root mean square velocity of the particle distribution, the resulting limit on $\Delta t$ will be appropriate for typical particles, but not all. In particular, particles in high-density regions will tend to have much larger velocities, in turn requiring smaller time steps. One could lower the proportionality factor of the \PTHREEM{} limiter accordingly, but at the cost of having unnecessarily fine time steps for the majority of the particles, wasting computational resources. Inspired by the approach of \citet{gadget2}, \CONCEPTONE{} instead allows each individual particle $i$ to be updated on a time scale $\Delta t/2^{\ell_i}$, where $\ell_i\in\mathbb{N}_0$ is called the \emph{rung}. Particles on rung 0 follows the global time-stepping, while particles on higher rungs receive short-range forces at a higher rate. The slowly varying and collectively computed long-range force remains as is, i.e.\ it follows the rhythm of rung 0.

With each particle assigned a rung, the system is evolved using a hierarchical scheme demonstrated by Figure~\ref{fig:timestepping}, here shown for $n_{\text{rung}}=3$ rungs. In practice, this number dynamically adapts as needed, though with a default maximum value of 8. Though particles act as `receivers' only during kicks of the given rung in which they are assigned, they act as `suppliers' for kicks within every rung. This asymmetry breaks strict symplecticity and momentum conservation, though the errors introduced are so small that this is of no concern\footnote{\GADGETFOUR{} \citep{gadget4} implements a time-stepping scheme similar to the one used in \CONCEPTONE{} as well as one with manifest momentum conservation. This other scheme does not deliver significant improvements to the accuracy, but does come at the cost of additional force computations.}.

To determine which rung $\ell_i$ a given particle $i$ belongs to, we impose that it must not accelerate across a certain fraction\footnote{In \GADGETTWO{} the corresponding parameter is called \texttt{ErrTolIntAccuracy} and typically has a value of $\eta = 0.025$, which is also chosen as the default value used by \CONCEPTONE{}.} $\eta$ of its softening length $\epsilon$ within the time $\Delta t/2^{\ell_i}$, disregarding its initial velocity. That is,
\begin{equation}
    \ell_i(t) = \max \left( 0,\, \left\lceil \log_2  \Delta t \sqrt{\frac{|\vec{a}_i(t)|}{2\eta\epsilon}} \right\rceil\right) \label{eq:rung}
\end{equation}
where $\vec{a}_i$ is the comoving acceleration proportional to $\dot{\vec{q}}_i$, which from \eqref{eq:eom} is $\dot{\vec{q}}_i / (a^2m)$. This is implemented as
\begin{equation}
    \vec{a}_i(t) = [\vec{q}_i(t) - \vec{q}_i(t_{\text{prev}})] \biggl[ m \int_{t_{\text{prev}}}^{t}\! \d t' a^2(t') \biggr]^{-1} \,,
\end{equation}
where $t_{\text{prev}} < t$ refers to the time of the previous short-range kick undertaken by the particle. At the beginning of the simulation no such previous time exists, and so a `fake' kick is computed without applying the resulting momentum updates.

\section{Code validation and comparison}\label{sec:validation_comparison}
This section seeks to demonstrate the correctness of the results obtained with \CONCEPTONE{}. This is done by comparing the power spectra of \CONCEPTONE{} simulations to those of similar \GADGET{} simulations, using both \GADGETTWO{} and \GADGETFOUR{}. This strategy thus presupposes the correctness of \GADGET{} itself, which is well motivated by its wide usage and thorough testing over the past two decades.

\subsection{Simulation setup}

\paragraph*{\GADGETHEADING{}-like \CONCEPTHEADING{} simulations}
A large effort has gone into making \CONCEPT{} consistent with general relativistic perturbation theory. Thus, the large-scale power spectrum obtained from \CONCEPT{} simulations is designed to agree with that of linear Einstein-Boltzmann codes such as \CLASS{} \citep*{class}, which is successfully demonstrated in \citet{concept_linnu,concept_dcdm,concept_de}. To this end, \CONCEPT{} makes use of the full \CLASS{} background and employs the $N$-body gauge \citep{Fidler:2015npa,Fidler:2016tir} framework. Initial conditions generated by \CONCEPT{} are thus in $N$-body gauge. During simulation, this gauge is preserved by continually applying linear gravitational effects from non-matter species\footnote{Here photons and neutrinos, both of which are necessarily part of the \CLASS{} cosmology.} to the particles, implemented using PM techniques.

This strategy of \CONCEPT{} for making simulations consistent with general relativistic perturbation theory (pioneered by \citet{cosira} with the \COSIRA{} code) is further adopted by the \PKDGRAVTHREE{} code \citep*{pkdgrav3,euclidemulator2}, though \GADGETFOUR{} remains purely Newtonian. For a proper comparison between \CONCEPT{} and \GADGET{}, we then need to run \CONCEPT{} in a `\GADGET{}-like' mode. We still generate all simulation initial conditions using \CONCEPT{} in its `standard' mode, and so the simulations start off in $N$-body gauge. This is contrasted with typical Newtonian setups, where the initial conditions are in no well-defined gauge at all, but has been back-scaled \citep{Fidler:2017ebh} from the full, linear $a=1$ solution in order to ensure agreement with relativistic perturbation theory on large scales at the present day. As we do not apply radiation perturbations during the simulations nor make use of back-scaled initial conditions, our simulations are not consistent with either approach. We stress that this does not affect the results in any appreciable way. What is important for the comparisons is that \CONCEPT{} and \GADGET{} makes use of exactly the same initial conditions and simulation approach.

Leaving out the general relativistic correction kicks during \CONCEPT{} evolution is easy, as these are only applied once explicitly specified in the parameter file. For the background evolution, \CONCEPT{} inherits the tabulated solution from \CLASS{} (incorporating radiation), whereas \GADGET{} solves the matter \textplus{} $\Lambda$ Friedmann equation internally. This simplified background can be used within \CONCEPT{} as well, in which case it is likewise solved internally by the $N$-body code. Lastly, the two codes differ in how they place the PM grid, \CONCEPTONE{} using cell-centred grid values and \GADGET{} using cell-vertex grid values. In effect, the PM grids of the codes are relatively displaced by half a grid cell, $L_{\varphi}/2$, in all three directions. This makes a difference as the positions of the particles in the initial conditions are specified with respect to absolute space, not the PM grid. Though any effect on results from a purely numerical aspect such as this goes to demonstrate non-convergence of the solution, it is preferable to use identical PM setups when the comparison is between codes, as opposed to the absolute result. Thus, for these tests, all grids within \CONCEPT{} (including that used for initial condition generation) has been switched to cell-vertex mode. With these changes to the standard \CONCEPT{} setup, we are ready to perform \GADGET{}-like \CONCEPT{} simulations\footnote{The \href{https://jmd-dk.github.io/concept/tutorial/gadget.html}{documentation} includes a section on how to perform \GADGET{}-like simulations in practice.}.

\begin{table}
\caption{Cosmological parameters used for all simulations.}
\label{tab:cosmo_params}
\begin{tabular}{cc}
    \hline
    Parameter              &  Value                          \\
    \hline
    $H_0$                  &  $\SI{67}{km.s^{-1}.Mpc^{-1}}$  \\
    $\Omega_{\text{b}}$    &  $0.049$                        \\
    $\Omega_{\text{cdm}}$  &  $0.27$                         \\
    $A_{\text{s}}$         &  $\num{2.1e-9}$                 \\
    $n_{\text{s}}$         &  $0.96$                         \\
    \hline
\end{tabular}

\bigskip

\caption{Simulation parameters used for all simulations unless explicitly stated otherwise, with the number of particles $N$ and the box size $L_{\text{box}}$ as free parameters.}
\label{tab:sim_params}
\begin{tabular}{lcc}
    \hline
    Parameter                            & Symbol               &  Value                              \\
    \hline
    Softening length                     &  $\epsilon$          &  $0.03\, L_{\text{box}}/\cbrt{N}$   \\
    PM grid size                         &  $n_{\varphi}$       &  $2\cbrt{N}$                     \\
    Short-/long-range force split scale  &  $x_{\text{s}}$      &  $1.25\, L_{\text{box}}/n_{\varphi}$  \\
    Short-range cut-off scale            &  $x_{\text{r}}$      &  $4.5\, x_{\text{s}}$                  \\
    Initial scale factor                 &  $a_{\text{begin}}$  &  $0.01$                             \\
    \hline
\end{tabular}

\bigskip

\caption{Simulation parameters specific to \GADGET{}.}
\label{tab:gadget_params}
\begin{tabular}{lc}
    \hline
    Parameter                           &  Value    \\
    \hline
    \texttt{MaxSizeTimestep}            &  $0.03$   \\
    \texttt{TypeOfOpeningCriterion}     &  $1$      \\
    \texttt{ErrTolForceAcc}             &  $0.005$  \\
    \texttt{TreeDomainUpdateFrequency}  &  $0.1$    \\
    \hline
\end{tabular}

\bigskip

\caption{Parameters for high-precision \GADGETTWO{} simulations.}
\label{tab:gadget_highprec}
\begin{tabular}{lc}
    \hline
    Parameter                           &  Value    \\
    \hline
    \texttt{ErrTolForceAcc}             &  $0.001$  \\
    \texttt{TreeDomainUpdateFrequency}  &  $0.05$   \\
    \hline
\end{tabular}
\end{table}

\begin{figure}
\includegraphics[width=\columnwidth]{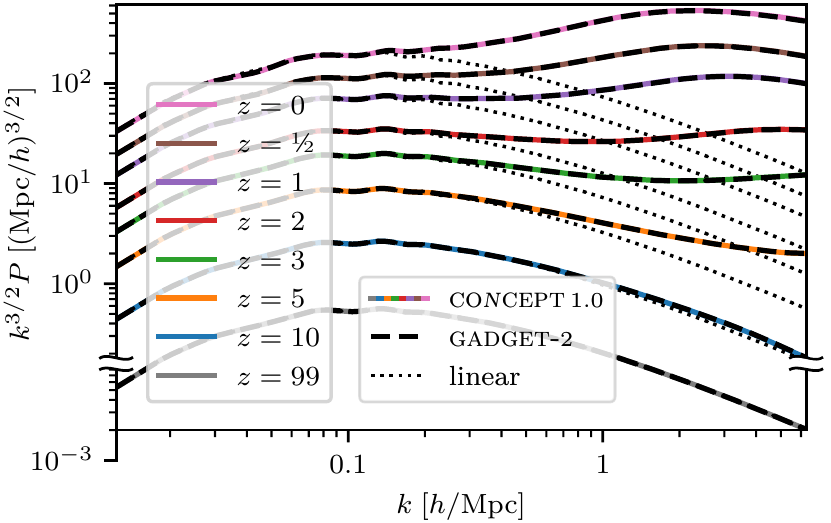}
\caption{Evolution of the power spectrum in \CONCEPTONE{} and \GADGETTWO{} simulations with $N=1024^3$ particles in a $\SI{512}{Mpc}/h$ box. For reference, the linear power spectrum is shown as well. Due to the large gap between the initial power at $z=99$ and the power at the first output time $z=10$, the vertical axis has been broken in two (with the entirety of the $z=99$ spectra belonging to the lower part). We plot $k^{3/2}P$ rather than $P$ or $k^3P$ as this results in less steep spectra, allowing for a more detailed view.
}
\label{fig:abspower}
\end{figure}

\paragraph*{Parameters}
All \CONCEPT{} and \GADGET{} simulations in this section use the cosmology specified in Table~\ref{tab:cosmo_params} and other simulation parameters as specified in Table~\ref{tab:sim_params}, with non-listed parameters taking on default\footnote{\CONCEPT{} inherits non-specified cosmological parameters from \CLASS{}.} \CONCEPTONE{} values. For \GADGET{} parameters that do not have a \CONCEPT{} equivalent, we likewise seek to employ default values. However, \GADGETTWO{} does not have a proper notion of default parameter values, and so we specify our chosen parameter values specific to \GADGETTWO{} in Table~\ref{tab:gadget_params}, with parameters not listed there (nor in Tables~\ref{tab:cosmo_params}~or~\ref{tab:sim_params}) taking on values as suggested by the \GADGETTWO{} user guide \citep{gadget2_userguide}. Parameters used with \GADGETFOUR{} likewise take on values as specified by Tables~\ref{tab:cosmo_params}, \ref{tab:sim_params}~and~\ref{tab:gadget_params} (when applicable), with parameters not listed taking on values as suggested by the \GADGETFOUR{} user manual \citep{gadget4_usermanual}. While we use \PTHREEM{} within \CONCEPTONE{}, we use TreePM within \GADGETTWO{} and various gravitational methods within \GADGETFOUR{}.

We settle for $N=1024^3$ particles and thus a PM grid of size $n_\varphi = 2048$, and run simulations for box sizes $L_{\text{box}} \in \{2048, 1024, 512, 256\}\,\si{Mpc}/h$. All power spectra are computed with \CONCEPT{} using a grid similarly of size $2048$, employing PCS interpolation \eqref{eq:W_PCS} and interlacing \citep{HockneyEastwood}.

\subsection{Comparison to \GADGETHEADING{}}

\begin{figure*}
\includegraphics[width=\textwidth]{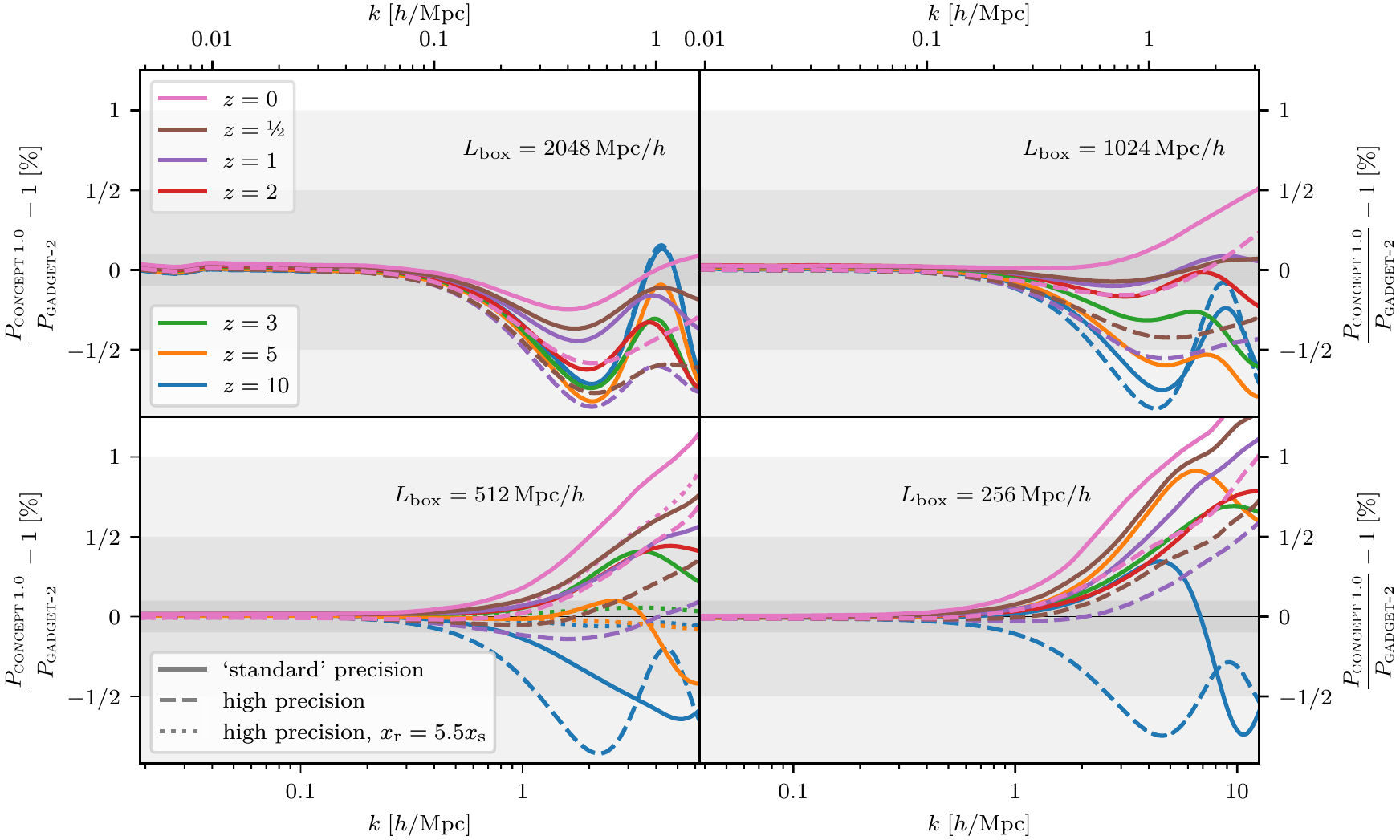}
\caption{
Relative power spectra between \CONCEPTONE{} and \GADGETTWO{}, for simulations with $N=1024^3$ particles and four different box sizes. The relative spectra are shown at various times, with the initial time $z=99$ left out as here the \CONCEPT{} and \GADGET{} spectra match exactly by construction. The full lines correspond to \GADGETTWO{} simulations using the `standard' \GADGETTWO{} precision settings of Table~\ref{tab:gadget_params}. For late times as well as $z=10$, the same \CONCEPT{} spectra are additionally shown relative to \GADGETTWO{} spectra from simulations utilising the high-precision settings of Table~\ref{tab:gadget_highprec}, using dashed lines. For the $\SI{512}{Mpc}/h$ box, we additionally show the case of increased $x_{\text{r}} = 5.5x_{\text{s}}$ in both \CONCEPTONE{} and \GADGETTWO{}, the latter again using the high-precision settings. These are shown as dotted lines and only for early times as well as $z=0$. Grey bands mark relative errors of $\SI{1}{\percent}$, $\SI[parse-numbers = false]{\text{\textonehalf}}{\percent}$ and $\SI{1}{\text{\textperthousand}}$. For each panel, the $k$ axis extends to the Nyquist scale of the particle grid, $k_{\text{Nyquist}} = \cbrt{N}/2 \times 2\PI/L_{\text{box}} = 1024\PI/L_{\text{box}}$.
}
\label{fig:correctness}
\end{figure*}

In Figure~\ref{fig:abspower} we show absolute power spectra from the \CONCEPT{} and \GADGETTWO{} simulation in the $\SI{512}{Mpc}/h$ box. Very good agreement between the codes is evident for all scales and times. This is impressive given that the non-linear power grows by more than a factor of $\num{2e+5}$ during the course of the simulations, and that the non-linear small-scale power at $z=0$ is more than $30$ times greater than its linear counterpart, demonstrating high non-linearity.

\paragraph*{\CONCEPTONEHEADING{} vs.\ \GADGETTWOHEADING{}}
For a more precise comparison between the \CONCEPT{} and \GADGETTWO{} results, their relative power spectra are shown in Figure~\ref{fig:correctness}, this time for all four box sizes. Here we see extraordinarily good agreement between the codes, for all scales and times irrespective of the box size. In all cases, the power spectra agree almost perfectly at large scales. Below some particular scale the results begin to diverge, with \CONCEPT{} predicting slightly less power than \GADGETTWO{} for large box sizes and early times (low clustering) and slightly more power than \GADGETTWO{} for small box sizes and late times (high clustering), culminating in $\sim\SI{1}{\percent}$ difference at the Nyquist scale.

Choosing a $\SI{1}{\text{\textperthousand}}$ relative difference as a proxy for the scale at which the results begin to diverge from each other, we find this scale to be $k_{\text{div}} \approx 24 \times 2\PI/L_{\text{box}}$, meaning it is relative to the resolution of the simulation(s) and does not depend on some absolute scale. That is, the difference between the codes is roughly independent on the box size / particle resolution. This $\SI{1}{\text{\textperthousand}}$ relative difference is shown in Figure~\ref{fig:correctness} as the innermost grey band.

We do see some difference as we vary the box size. In particular, \CONCEPT{} predicts slightly less power than \GADGETTWO{} for large boxes and slightly more power than \GADGETTWO{} for small boxes. As the main difference between the codes is that \GADGETTWO{} approximates the short-range force using a tree while \CONCEPT{} does not, we might hope that this difference is the main source of their disagreement. To test this we additionally run \GADGETTWO{} using higher-precision tree settings as listed in Table~\ref{tab:gadget_highprec} (all other parameters stay the same), traversing the tree more deeply and rebuilding it anew more frequently. The results of such high-precision \GADGETTWO{} simulations are also shown in Figure~\ref{fig:correctness}, compared against the same \CONCEPT{} results as before. For all boxes, increasing the tree precision of \GADGETTWO{} has the effect of lowering the power, leading to better agreement with \CONCEPT{} for the smaller box sizes. Interestingly, improving the tree approximation worsens the agreement for the larger box sizes and for early times generally. This is most likely related to the ``fuzzy short-range interaction boundary'' of \GADGETTWO{} discussed further down.

Increasing the precision of the tree force as in Table~\ref{tab:gadget_highprec} has another effect. Looking carefully at all but the smallest box of Figure~\ref{fig:correctness}, we see that the large-scale \CONCEPTONE{} power very slightly disagree (at a few tens of a per mille) with that of the `standard'-precision \GADGETTWO{} simulations, whereas this constant offset drops by a factor $\sim 10$ with the high-precision \GADGETTWO{} runs. We stress that even with the `standard'-precision \GADGET{} runs, this constant offset is very tiny. Indeed, in order to obtain this good of an agreement, we have had to update the values of various physical constants used in \GADGETTWO{} to match the exact values used in \CONCEPTONE{}. Here the most important one is probably the gravitational constant, which \GADGETTWO{} sets to $G = \SI{6.672e-11}{m^3.kg^{-1}.s^{-2}}$ whereas \CONCEPTONE{} uses the latest value from the \citet{pdg2020} $G = \SI{6.67430e-11}{m^3.kg^{-1}.s^{-2}}$. Without this matching of the values of physical constants, the constant offset between \GADGETTWO{} and \CONCEPTONE{} grows by a factor $\sim 2.5$, though it stays below one per mille.

The relative spectra at the largest scales for the largest box size $L_{\text{box}} = \SI{2048}{Mpc}/h$ develops a slight but persistent wiggle early on. This effect not only remains but worsens for still larger boxes, and so it is associated with large physical scales, irrespective of the simulation resolution. The feature is robust against increased temporal precision of either code, and also against lowering the tree opening angle within \GADGETTWO{}. However, the wiggle can be made to completely disappear by running \GADGETTWO{} with a slightly increased short-range cut-off scale, $x_{\text{r}} \gtrsim 5.0\,  x_{\text{s}}$. This is surprising, as the short-range force should have no effect on the largest scales. Indeed, running \CONCEPT{} with a similarly increased $x_{\text{r}}$ only perturbs its spectrum at small scales, leaving the larger scales invariant.

While Figure~\ref{fig:correctness} does not show the case of increased $x_{\text{r}}$ for the largest box size $L_{\text{box}} = \SI{2048}{Mpc}/h$, it does show $x_{\text{r}}=5.5x_{\text{s}}$ for $L_{\text{box}} = \SI{512}{Mpc}/h$ at a few selected times. Here both codes are run with this increased $x_{\text{r}}$ and \GADGETTWO{} is further run using the high-precision settings. At early times, we see that this slight increase to $x_{\text{r}}$ reduces the discrepancy between \CONCEPTONE{} and \GADGETTWO{} to the point where they now agree at the per mille level at all relevant scales. We believe this to be explained by the cut-off $x_{\text{r}}$ being strictly enforced at the particle-particle level in \CONCEPT{}, whereas the tree in \GADGETTWO{} makes this cut-off somewhat fuzzy due to the physical extent of its nodes. At low clustering this difference will be particularly pronounced as a lot of precise force-cancellation takes place for the near-homogeneous particle distribution. In Figure~\ref{fig:correctness} we indeed only find a deficit of power in \GADGETTWO{} relative to \CONCEPTONE{} at low clustering (large boxes and/or early times). Increasing $x_{\text{r}}$ pushes the fuzzy interaction boundary in \GADGETTWO{} to greater particle separations, with the short-range force exponentially decaying, decreasing its significance. Interestingly, while increasing $x_{\text{r}}$ leads to better agreement at early times, Figure~\ref{fig:correctness} also shows that it in fact slightly worsens the agreement at $z=0$.

\begin{figure}
\includegraphics[width=\columnwidth]{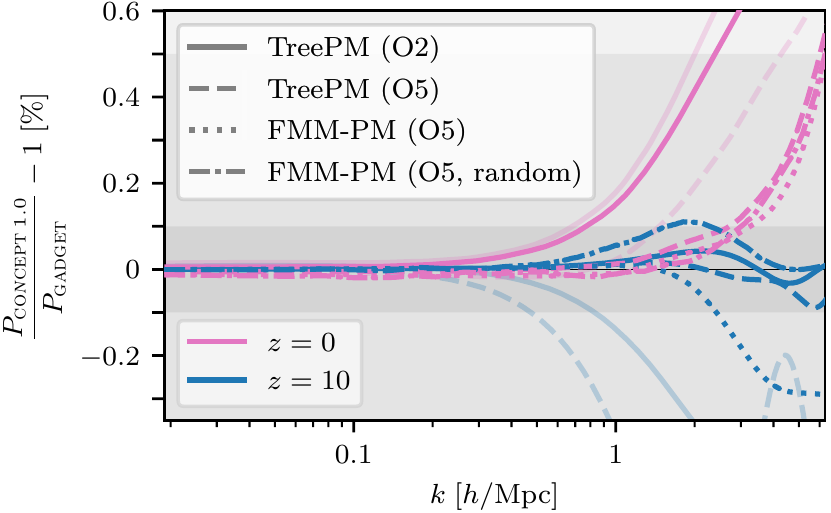}
\caption{
Relative power spectra between \CONCEPTONE{} and \GADGETFOUR{}, for simulations with $N=1024^3$ particles in a box of size $L_{\text{box}} = \SI{512}{Mpc}/h$, at early $z=10$ and late $z=0$ times. The full and dashed lines show the case of \GADGETFOUR{} using the TreePM method, with multipole expansion order 2 and 5, respectively. The dotted and dash-dotted lines show the case of \GADGETFOUR{} using the FMM-PM method with multipole expansion order 5, where the dash-dotted in addition use randomised displacements. All \GADGETFOUR{} simulations use the `standard' \GADGET{} settings of Table~\ref{tab:gadget_params}. This figure is similar to the lower left panel of Figure~\ref{fig:correctness}, where \CONCEPTONE{} is compared against \GADGETTWO{}. In fact, the shaded lines show the `standard'- and high-precision \GADGETTWO{} data (full and dashed, respectively) from Figure~\ref{fig:correctness}, for reference. Grey bands mark relative errors of $\SI{5}{\text{\textperthousand}}$ and $\SI{1}{\text{\textperthousand}}$. The $k$ axis extends to the Nyquist scale of the particle grid, $k_{\text{Nyquist}} = \cbrt{N}/2 \times 2\PI/L_{\text{box}} = \SI[parse-numbers = false]{2\PI}{Mpc}/h$.
}
\label{fig:correctness_512}
\end{figure}

\paragraph*{\CONCEPTONEHEADING{} vs.\ \GADGETFOURHEADING{}}
We have seen that the \CONCEPTONE{} and \GADGETTWO{} codes generally agree very well, with even better late-time agreement obtainable by increasing the tree precision parameters of \GADGETTWO{}. For better agreement at early times, we had to not just use these so-called high-precision \GADGET{} settings, but also increase the short-range cut-off scale $x_{\text{r}}$ within both codes. With these findings in mind, let us now compare the results of \CONCEPTONE{} to those of \GADGETFOUR{}.

Figure~\ref{fig:correctness_512} shows relative power spectra between \CONCEPTONE{} and \GADGETFOUR{}, for simulations with $N = 1024^3$ particles in a box of size $L_{\text{box}} = \SI{512}{Mpc}/h$, i.e.\ it is similar to the lower left panel of Figure~\ref{fig:correctness}, though with \GADGETTWO{} substituted with \GADGETFOUR{}. The results of four different gravitational methods used within \GADGETFOUR{} are shown.

One of the gravitational methods used by \GADGETFOUR{} as shown in Figure~\ref{fig:correctness_512} is that of TreePM with multipole expansion order 2. This is also the gravitational method used by all our \GADGETTWO{} simulations (this multipole order is fixed in \GADGETTWO{}), and so we might expect these \GADGETFOUR{} results to closely match those of our `standard'  \GADGETTWO{} runs. At late times $z=0$, we indeed obtain results comparable to what we got with \GADGETTWO{}, though now with slight improved agreement with \CONCEPTONE{}. However, at early times $z=10$, \GADGETFOUR{} delivers results which agree much better with \CONCEPTONE{} than those of \GADGETTWO{}, rivalling even the results obtained with high-precision \GADGETTWO{} and increased $x_{\text{r}}$.

In \GADGETFOUR{} the multipole expansion order is adjustable, allowing us to improve on the approximation to the short-range tree-force. As \CONCEPTONE{} does not approximate this force at all (it has no tree), we expect higher orders to improve the agreement with \CONCEPTONE{}. While the order-5 TreePM results on Figure~\ref{fig:correctness_512} does not show significant difference compared to TreePM order-2 at early times $z=10$, it yields a significant improvement at late times $z=0$. In fact, `standard'-precision \GADGETFOUR{} with order-5 TreePM agrees even better with \CONCEPTONE{} than does \emph{high}-precision \GADGETTWO{}, with the size of the additional improvement comparable to the difference between `standard'- and high-precision \GADGETTWO{}.

Besides TreePM, \GADGETFOUR{} further implements FMM-PM \citep{gadget4}, which retains the usual long-range particle-mesh force but supply the short-range force using a Fast Multipole Method (FMM). The FMM \citep{fmm} replaces the particle-node interaction of the usual one-sided tree \citep{Barnes:1986nb} with a symmetric node-node interaction, allowing for manifest momentum conservation. As the (sub)tile-(sub)tile interaction of the short-range force within \CONCEPTONE{} is always resolved completely to the particle-particle level, gravity in \CONCEPTONE{} is similarly momentum-conserving. Figure~\ref{fig:correctness_512} includes an order-5 \GADGETFOUR{} FMM-PM run, which at late times $z=0$ looks similar to the order-5 Tree-PM run. The early time $z=10$ behaviour is however noticeably different, with a several per mille drop in power at high $k$.

Figure~\ref{fig:correctness_512} further shows another order-5 FMM-PM \GADGETFOUR{} run, this time with randomised displacements enabled, effectively shifting the tree and grid geometries relative to the particle distribution by a random offset at each time step. This reduces the temporal correlation of force errors, as otherwise slowly moving particles will receive the same force error over many time steps, due to them being situated in more or less the same spot relative to the tree nodes and the force grid cells, the geometries of which affect the force in a non-physical manner. Though such correlations of force errors exist for all the gravitational methods used, they are particularly harmful for FMM, as demonstrated in \cite{gadget4}. Indeed, as evident from Figure~\ref{fig:correctness_512}, enabling randomised displacements fixes the early $z=10$ behaviour of order-5 FMM-PM, while leaving the late $z=0$ behaviour practically the same.

Unlike the tree in \GADGETTWOFOUR{} (be it that of TreePM or FMM-PM), the (sub)tile geometry of \CONCEPTONE{} has no effect on the final short-range force felt by a particle\footnote{Up to very small differences arising from the non-associativity of floating-point addition.}, and so it is reasonable that better agreement is obtained when making an effort to reduce force correlation errors inside \GADGETFOUR{}. As \CONCEPTONE{} does have a potential grid, some correlation errors are still expected\footnote{One way to reduce correlation errors from the potential grid in \CONCEPTONE{} is to enable interlacing \citep{HockneyEastwood} for the interpolating of particles onto the grid, which shrinks the effective grid volume by a factor of 8.}. As the grid cells in use are physically rather small, $L_\varphi = L_{\text{box}}/(2\cbrt{N}) = \SI{0.25}{Mpc}/h$, we might expect the correlation errors from the grid to be much smaller than those from the tree, matching the observation that running \GADGETFOUR{} with randomised displacements only seem to improve the agreement between it and \CONCEPTONE{}.

Figure~\ref{fig:correctness_512} tells us that \GADGETFOUR{} produces results even more similar to those of \CONCEPTONE{} than does \GADGETTWO{}, primarily due to the availability of higher-order multipole expansions. At least for such higher-order multipole expansions, whether running \GADGETFOUR{} with TreePM or FMM-PM gravity does not change the results appreciably. While increasing the precision of the tree in \GADGETTWO{} in accordance with Table~\ref{tab:gadget_highprec} brings the results significantly closer to those of \GADGETFOUR{} and \CONCEPTONE{}, we have observed no significant change from a similar\footnote{While the \texttt{ErrTolForceAcc} parameter is retained in \GADGETFOUR{}, the \texttt{TreeDomainUpdateFrequency} parameter is no longer available.} increase to the precision of the tree in \GADGETFOUR{}.

\section{Code performance}\label{sec:performance}
With the correctness of \CONCEPTONE{} established by the previous section, we now set out to demonstrate various performance aspects of the code, both internally and by comparison to \GADGETTWOFOUR{}. All simulations employed in this section use the cosmology as specified in Table~\ref{tab:cosmo_params} along with other simulation parameters as specified in Tables~\ref{tab:sim_params}~and~\ref{tab:gadget_params}, as in the previous section.

All simulations (of this and the previous section) are carried out on the Grendel compute cluster at Centre for Scientific Computing Aarhus (CSCAA), using compute nodes each consisting of two 24-core Intel Xeon Gold 6248R CPUs at $\SI{3.0}{GHz}$, interconnected with Mellanox EDR Infiniband at $\SI{100}{Gbit/s}$. All cores have hyper-threading disabled. Both \CONCEPTONE{} and \GADGETTWOFOUR{} are built using GCC 10.1.0 with optimisations \texttt{-O3 -funroll-loops -ffast-math -flto} and linked against FFTW 3.3.9 (\CONCEPTONE{} and \GADGETFOUR{}) or 2.1.5 (\GADGETTWO{}), itself built similarly though without link-time optimisations \texttt{-flto}. All of \CONCEPTONE{}, \GADGETTWOFOUR{} and FFTW 2/3 are run in double-precision. All is linked against and run with OpenMPI 4.0.3. Specifically, we use version 1.0.0 of \CONCEPT{}, version 2.0.7 of \GADGETTWO{} and Git commit \texttt{8a10478b3e62d202808407e40a5f94a8b5e88d80} of \GADGETFOUR{}.

\subsection{Weak scaling}
Here we study the `weak scaling' of \CONCEPTONE{}, i.e.\ how the computation time is affected for increased problem size while keeping the computational load per process fixed. That is, we hold $L_{\text{box}}\propto \cbrt{N}$ and $n_{\text{p}}\propto N$ for varying $N$, with $n_{\text{p}}$ being the total number of MPI processes, each running on a dedicated CPU core. For perfect weak scaling, increasing the problem size together with the number of CPU cores in lockstep should not incur any increase to the computation time.

\begin{figure*}
\includegraphics[width=\textwidth]{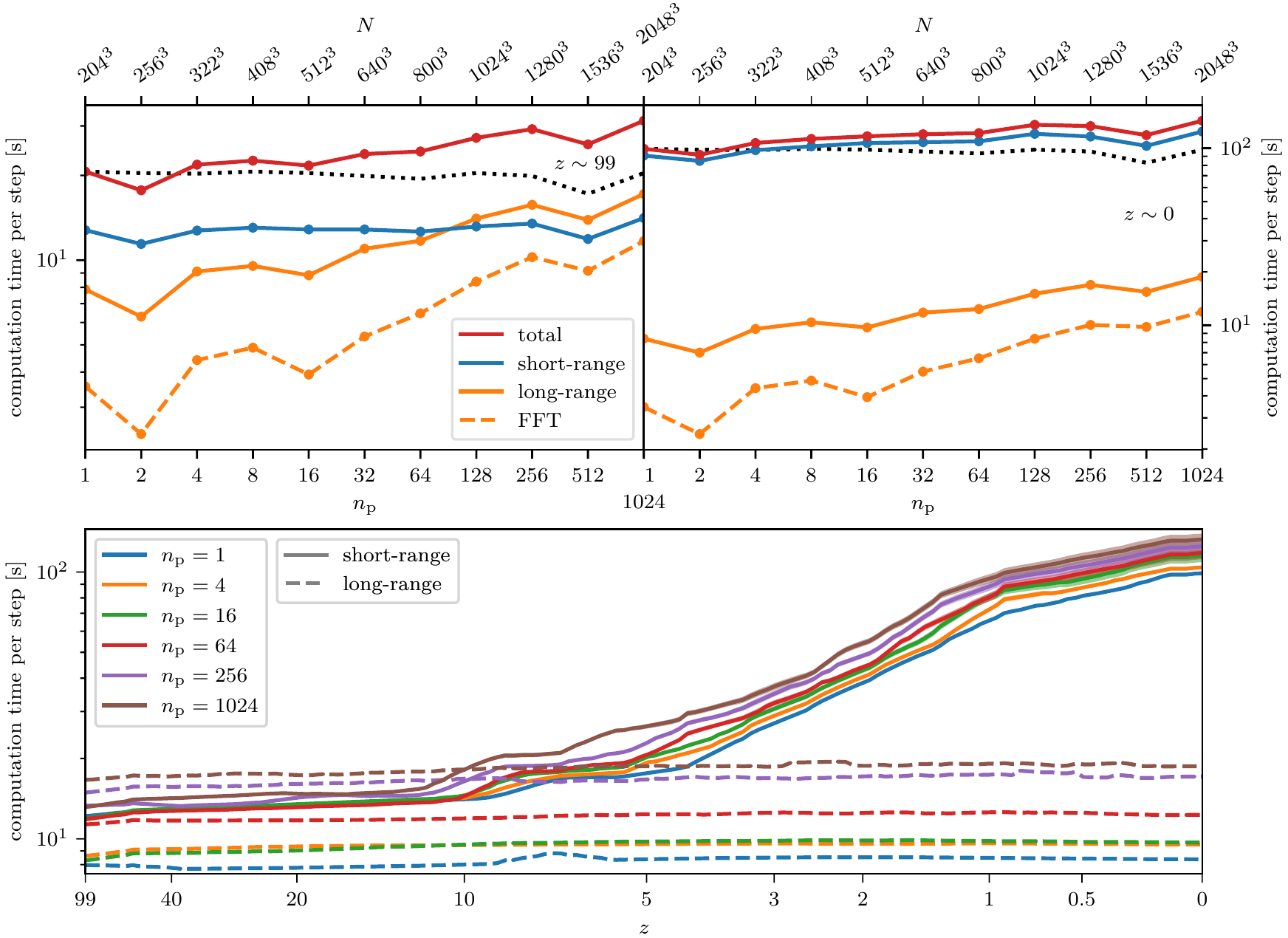}
\caption{
Weak scalability for \CONCEPTONE{} simulations in boxes of size $L_{\text{box}} = \SI[parse-numbers = false]{2\cbrt{N}}{Mpc}/h$, keeping the particle load per process $N/n_{\text{p}}$ roughly fixed at $204^3$ particles. The simulations range from serial all the way to $1024$ cores and $2048^3$ particles. The top panels show the wall-clock computation time per time step near the beginning $z\sim 99$ and end $z\sim 0$ of the simulations, averaged over 8 time steps. Full lines show the total computation time and its dominant components, namely the short-range and long-range gravitational computations. The FFT part of the long-range computation is further separated out and shown using dashed lines. Finally, perfect weak scaling of the total computation time is shown with dotted lines. We cannot explain the dip at $n_{\text{p}}=2$, which is seen in both the short- and long-range computation time. The dip at $n_{\text{p}}=512$ is caused by having slightly smaller particle load per process than usual (note that this dip appears even in the perfect scaling). Here we ought to use $N\approx 1632^3$, but $n_{\varphi}=2\sqrt[2]{N}$ must be divisible by $n_{\text{p}}=512$ due to restrictions in \CONCEPT{} (the FFTW slabs must be evenly divisible amongst the processes).\newline
The lower panel shows the evolution of the computation time over the simulation time span for every other simulation, averaged over 8 time steps. Here only the short-range (full) and long-range (dashed) computation times are shown. Towards $z=0$ the load imbalance can be seen as a widening of the short-range lines, with the widths given by twice the standard deviation of the individual short-range computation times among the processes within a given simulation. The redshift $z$ axis is shown as scaling linearly with the simulation time steps.
}
\label{fig:weak}
\end{figure*}

Choosing $L_{\text{box}} = \SI[parse-numbers = false]{2\cbrt{N}}{Mpc}/h$ and $N/n_{\text{p}} \sim 204^3$, the weak scaling of \CONCEPTONE{} is shown in Figure~\ref{fig:weak}. From the top panels, we see that the short-range computation exhibits almost perfect weak scaling at early times and still reasonably good weak scaling at late times. The long-range computation has a less optimal scaling, even overtaking the short-range computation at early times when having many processes. This suboptimal scaling of the long-range computation is owed mostly to the FFTs, as evident from the dashed orange lines having similar shape to the full orange lines but with steeper slope. At late times the computation time is completely dominated by the short-range computation, rendering the bad scaling of the long-range part ignorable. In all, this leads to reasonably good overall weak scaling of \CONCEPTONE{}.

Looking at the lower panel of Figure~\ref{fig:weak}, the suboptimal weak scaling of the long-range computation is again evident, here as clear separations between (most of) the dashed lines. The long-range computation time is however close to constant throughout the simulation. For the short-range times, the different simulations follow each other closely, though still with larger simulations being somewhat slower. The cost of the short-range computation increases as the universe becomes more clustered. Here this effect kicks in at $z\sim 10$ and continues to the present day. This increase is caused by the particle-particle interaction count going up with the amount of clustering. As the load imbalance remains small even at late times and high core count, this is not a significant factor in the slowdown of the short-range computation over the course of the simulation time span.

\subsection{Strong scaling}

\begin{figure*}
\includegraphics[width=\textwidth]{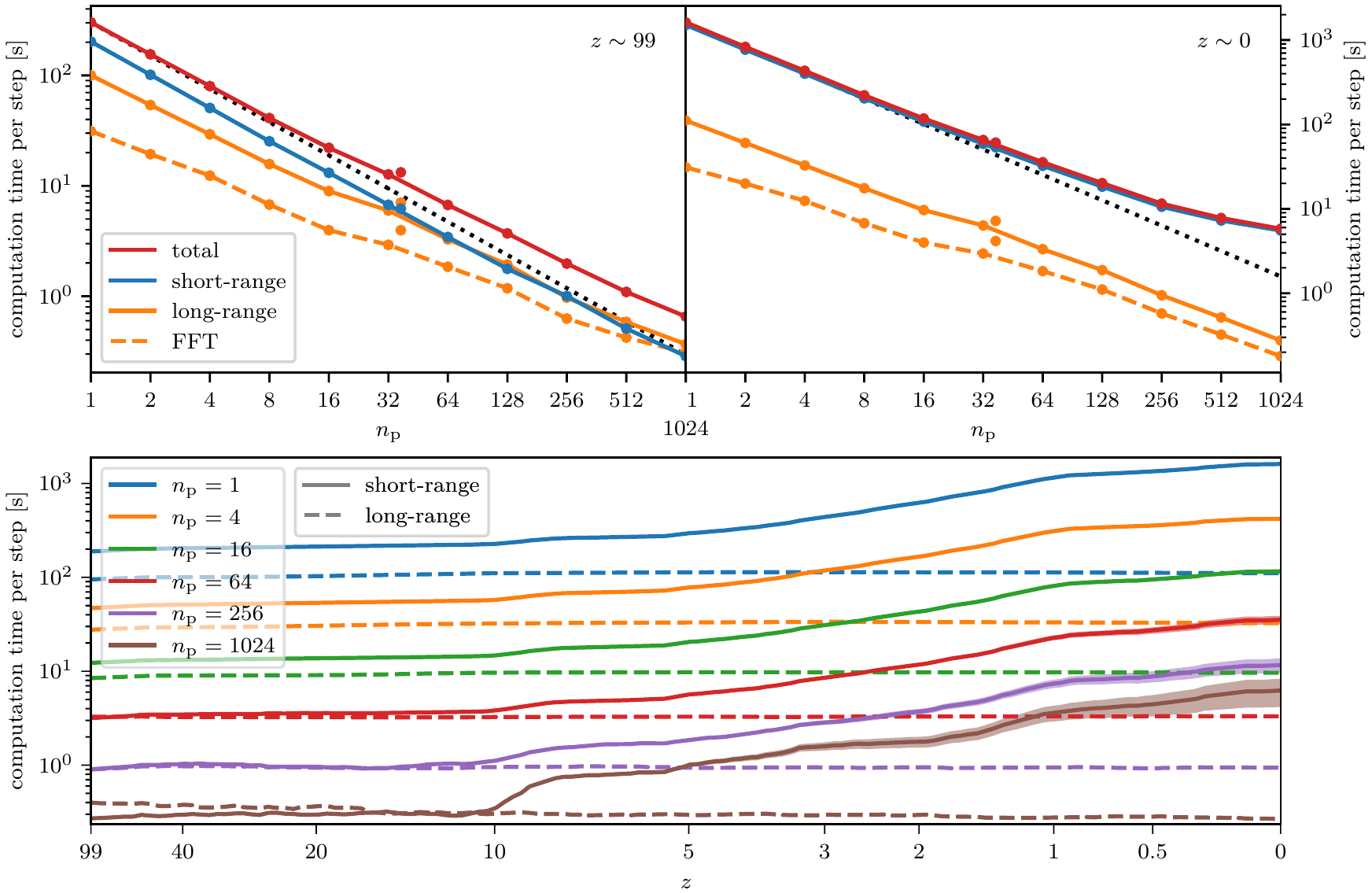}
\caption{
Strong scalability for \CONCEPTONE{} simulations with $N = 512^3$ particles in a box of size $L_{\text{box}} = \SI{1024}{Mpc}/h$. The simulations range from serial all the way to $1024$ cores, corresponding to a load ranging from $512^3$ to $\sim 50^3$ particles per core. The top panels show the wall-clock computation time per time step near the beginning $z\sim 99$ and end $z\sim 0$ of the simulations, averaged over 8 time steps. Full lines show the total computation time and its dominant components, namely the short-range and long-range gravitational computations. The FFT part of the long-range computation is further separated out and shown using dashed lines. Finally, perfect strong scaling of the total computation time is shown with dotted lines. In addition to having the number of processes being powers of two, we further show the case of $n_{\text{p}}=37$ as disconnected data points. As \CONCEPT{} requires $n_{\varphi}$ to be divisible by $n_{\text{p}}$, this one simulation has been run with slightly increased grid size $n_{\varphi} = 1036$ instead of the usual $n_{\varphi} = 2\cbrt{N} = 1024$ used for the other simulations.\newline
The lower panel shows the evolution of the computation time over the simulation time span for every other simulation, averaged over 8 time steps. Here only the short-range (full) and long-range (dashed) computation times are shown. Towards $z=0$ the load imbalance can be seen as a widening of the short-range lines, with the widths given by twice the standard deviation of the individual short-range computation times among the processes within a given simulation. The redshift $z$ axis is shown as scaling linearly with the simulation time steps.
}
\label{fig:strong}
\end{figure*}

Here we study the `strong scaling' of \CONCEPTONE{}, i.e.\ how the computation time is affected when increasing the number of CPU cores used for the simulation, keeping everything else fixed. That is, for some chosen $L_{\text{box}}$ and $N$ we vary $n_{\text{p}}$. For perfect strong scaling, the computation time is required to drop linearly with the number of cores, i.e.\ the computation time should be inversely proportional to the computational firepower thrown at the problem.

Figure~\ref{fig:strong} shows the strong scaling of \CONCEPTONE{} for $L_{\text{box}} = \SI{1024}{Mpc}/h$, $N = 512^3$. The short-range computation scales very well, especially at early times, as evident from the upper panels of the figure. The long-range computation shows a somewhat worse strong scaling behaviour than the short-range computation, even overtaking as the dominant computation for high core counts at early times. As evident from the similar shape of the full and dashed orange curves, this behaviour is due to the FFTs.

The sudden jump in the trend line of the long-range computation time at $n_{\text{p}}\geq 32$ is probably explained by the $n_{\text{p}} < 32$ simulations all running entirely within a single CPU, whereas the $n_{\text{p}}\geq 32$ simulations all utilise several CPUs, even distributed over several compute nodes for $n_{\text{p}}\geq 64$. As the short-range computation vastly dominates at late times, this suboptimal strong scaling of the long-range force is not an issue in practice. In total, this makes the overall strong scaling of \CONCEPTONE{} reasonably good.

The top panels of Figure~\ref{fig:strong} include the odd case of $n_{\text{p}}=37$, a prime. This is to demonstrate that \CONCEPTONE{} may be run with any number of processes and that the nature of this number does not significantly affect its performance. The computation time of the FFTs does increase noticeably, but as usual this effect is dwarfed by the dominance of the short-range computation at late times. Over the course of a whole simulation then, the nature of $n_{\text{p}}$ is of little importance.

For the lower panel of Figure~\ref{fig:strong}, decent strong scalings of the short- and long-range computations are evident from the nearly equidistant separations between the lines. As for the weak scaling results of Figure~\ref{fig:weak}, we again find the computation time of the long-range force to be mostly invariant over the simulation time span, and that the cost of the short-range computation increases as the universe becomes more clustered. At late times, load imbalance starts to become significant for the simulations with large core counts, degrading the strong scaling.

\subsection{Absolute performance}
The above explorations of the weak and strong scaling of \CONCEPTONE{} demonstrate excellent scaling behaviour when increasing the problem size $N$ and/or the core count $n_{\text{p}}$. Keeping both of these fixed, the computation time required for a given simulation depends on the level of clustering, which in turn depends on the particle resolution through the box size $L_{\text{box}}$. We now want to investigate the absolute performance of \CONCEPTONE{} as a function of the particle resolution, which we do by comparing the total computation time of \CONCEPTONE{} simulations to equivalent \GADGETTWO{} simulations.

Even though Figure~\ref{fig:correctness} generally demonstrates improved agreement between \CONCEPTONE{} and \GADGETTWO{} for the high-precision \GADGET{} settings of Table~\ref{tab:gadget_highprec}, we here exclusively run \GADGETTWO{} with the `standard' settings of Table~\ref{tab:gadget_params}. We choose to do so as it would be unfair not to allow \GADGETTWO{} to make good use of its tree approximation when comparing performance, given that the observed improvements brought about by the high-precision settings are relatively minor. For the two larger boxes of Figure~\ref{fig:correctness}, running \GADGETTWO{} with the high precision settings only incurs a performance hit of a few percent, though this grows to $\sim\SI{30}{\percent}$ for the two smaller boxes. Likewise, despite the improved agreement with \CONCEPTONE{} observed in Figure~\ref{fig:correctness_512}, we do not include \GADGETFOUR{} in this performance comparison, as we have found \GADGETFOUR{} to be about two (three) times as slow as \GADGETTWO{} when using TreePM (FMM-PM), almost independent of multipole expansion order and whether or not randomised displacements are in use.

\begin{figure}
\includegraphics[width=\columnwidth]{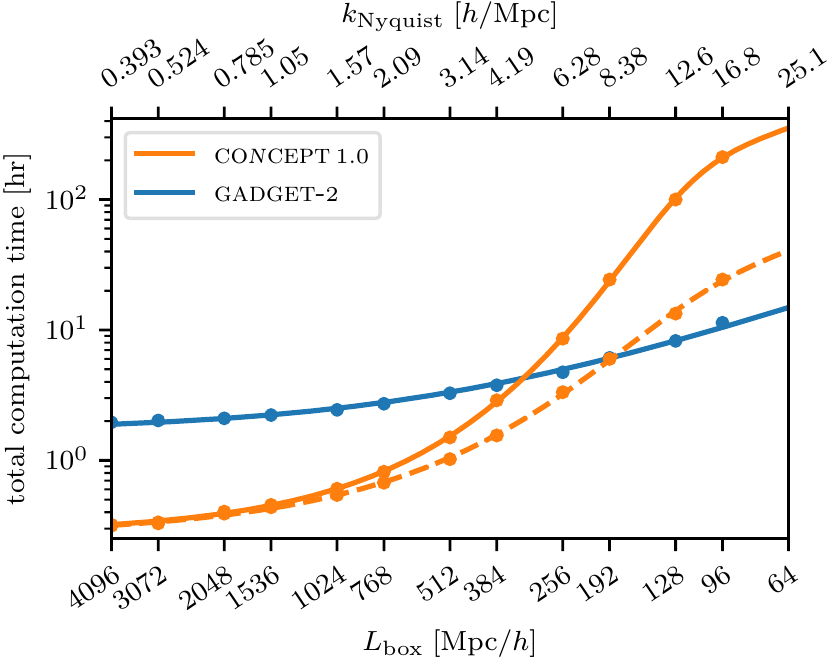}
\caption{Scalability across particle resolutions for \CONCEPTONE{} and \GADGETTWO{} simulations with $N=512^3$ particles, all run using 64 processes evenly distributed across two dedicated compute nodes. The wall-clock computation times of entire simulations are plotted against the box size $L_{\text{box}}$, from a very large box $L_{\text{box}} = \SI{4096}{Mpc}/h$ down to a very small box $L_{\text{box}} = \SI{96}{Mpc}/h$. Alternatively, the horizontal axis may be viewed in terms of the Nyquist scale of the particle grid, $k_{\text{Nyquist}} = \cbrt{N}/2 \times 2\PI/L_{\text{box}} = 512\PI/L_{\text{box}}$. The blue line is a linear (in $k_{\text{Nyquist}}$) fit to the \GADGETTWO{} data points, whereas the full orange line is a fit to the \CONCEPTONE{} data points using separate linear behaviour at either end and a sigmoid transition between the two. The dashed orange line is constructed from the full orange line by subtracting wasted short-range computation time due to load imbalance, and so represent the absolute performance of \CONCEPTONE{} had it contained a perfect load balancing scheme.
}
\label{fig:box}
\end{figure}

In Figure~\ref{fig:box} we plot the total computation times of \CONCEPTONE{} and \GADGETTWO{} simulations for various box sizes, corresponding to Nyquist scales of the particle grid ranging from $k_{\text{Nyquist}} \sim 0.4\, h/\si{Mpc}$ to $k_{\text{Nyquist}} \sim 17\, h/\si{Mpc}$. For $k_{\text{Nyquist}} \lesssim 5\,\si{Mpc}/h$ \CONCEPTONE{} is much faster than \GADGETTWO{}, whereas \GADGETTWO{} is much faster than \CONCEPTONE{} for $k_{\text{Nyquist}} \gtrsim 5\,\si{Mpc}/h$.

We believe that the superior performance of \CONCEPTONE{} at low to moderate clustering has two primary causes. First, the non-hierarchical tile \textplus{} subtile data structure of \CONCEPTONE{} is much faster to traverse than the tree structure of \GADGET{}, due to simple, precomputed access patterns and minimal pointer chasing. At low clustering, all particles have a similar number of short-range interaction partner particles, and so the benefits of the grouping carried out by the tree is minimal. At stronger clustering, the number of particle-particle short-range interactions increases drastically, which is then efficiently approximated by much fewer particle-node interactions using the tree, outweighing the more expensive tree walk. Second, \CONCEPTONE{} employs a much coarser time-stepping at high redshift than \GADGET{}, as discussed in section~\ref{subsec:timestepping}. As evident from Figure~\ref{fig:correctness} this does not induce noticeable artefacts in the solution.

The slowness of \CONCEPTONE{} at very high resolution means that it is currently impractical to use the code for simulations in this regime. Though a tree implementation in \CONCEPT{} would undoubtedly speed up the expensive short-range computation at these resolutions, Figure~\ref{fig:correctness} reveals a more important possible optimisation; load balancing. Currently \CONCEPT{} does no attempt at balancing the computational load across the CPU cores, as discussed in section~\ref{subsubsec:p3m_gravity}. For large clustering, this leads to correspondingly large load imbalance of the short-range computation, as visible in e.g.\ the lower panel of Figure~\ref{fig:strong}. The dashed line in Figure~\ref{fig:box} shows the theoretical computation time of \CONCEPTONE{} runs with the load perfectly balanced (assuming the balancing itself is cost free), and as so represents the best performance improvement we can hope to obtain were we to build load balancing into \CONCEPT{}. Though still slower than \GADGETTWO{} for runs with very high resolution, this alone would be enough to make it feasible to perform such simulations with \CONCEPT{}.

The data points of Figure~\ref{fig:box} are fitted to trend-lines. In the case of \GADGETTWO{}, a simple linear fit match the data nicely. In the case of \CONCEPTONE{}, the scaling behaviour is less trivial. In the low-resolution regime \CONCEPTONE{} exhibits linear scaling as well. The other extreme is trickier to gauge due to scarcity of data, but the fits suggests that here too it moves towards (a different) linear scaling, both in the actual case and with perfect load balancing. The different scaling behaviours at the two ends reflect the fact that at high resolution the short-range force completely dominates the computational budget, whereas at low resolution the long-range computation is comparably (if not more) expensive. In the case of \GADGET{}, both the short- and long-range computation scales as $\mathcal{O}(N \log N)$, and so shifting the computational burden from one to the other does not significantly change the scaling behaviour.

\subsection{Memory consumption}
With the previous subsections having thoroughly investigated the time complexity of \CONCEPTONE{}, let us now turn to its space complexity (consumption of memory).

To understand the memory usage of \CONCEPTONE{}, we simply tally up\footnote{In doing so, we assume a standard x86-64 architecture.} the memory consumed by its major data structures, most important of which are the particle data arrays and the \PTHREEMPAREN{} mesh\footnote{Here and later we use `\PTHREEMPAREN{}' to mean `PM or \PTHREEM{}'.}. The canonical vector variables of each particle contribute to the memory budget with 3 triplets of 8-byte (i.e.\ double-precision) floating-point numbers (position $\vec{x}_i$, momentum $\vec{q}_i$, momentum update $\Delta\vec{q}_i$), as well as 3 1-byte integers for keeping track of the rung ${\ell}_i$. The tiling brings in another 8-byte integer per particle. At late times the number of allocated particles somewhat exceeds $N$ due to particle exchange between the processes. The memory spent on the particles thus slightly increases during the simulation, and so the above memory consumption should be scaled up by some small factor, say $\sim 1.25$.

Each of the $n^3_{\varphi}$ \PTHREEMPAREN{} grid cells store an 8-byte floating-point number, with 3 such global grids present in memory (domain-decomposed potential, slab-decomposed potential, force). Altogether, this yields a memory consumption of $M \approx (104\, N + 24\, n_{\varphi}^3)\,\si{B}$, where $\si{B}$ is a byte. The tiles and their pre-computed pairings further contribute noticeably to the total memory, as do various buffers. Aided by measurements, we find the true memory consumption to be closer to
\begin{equation}
    M \approx (120\, N + 28.3\, n_{\varphi}^3)\,\si{B}\,, \label{eq:mem_usage}
\end{equation}
given the \PTHREEM{} parameters~\eqref{eq:p3m_parameters}.

\begin{figure}
\includegraphics[width=\columnwidth]{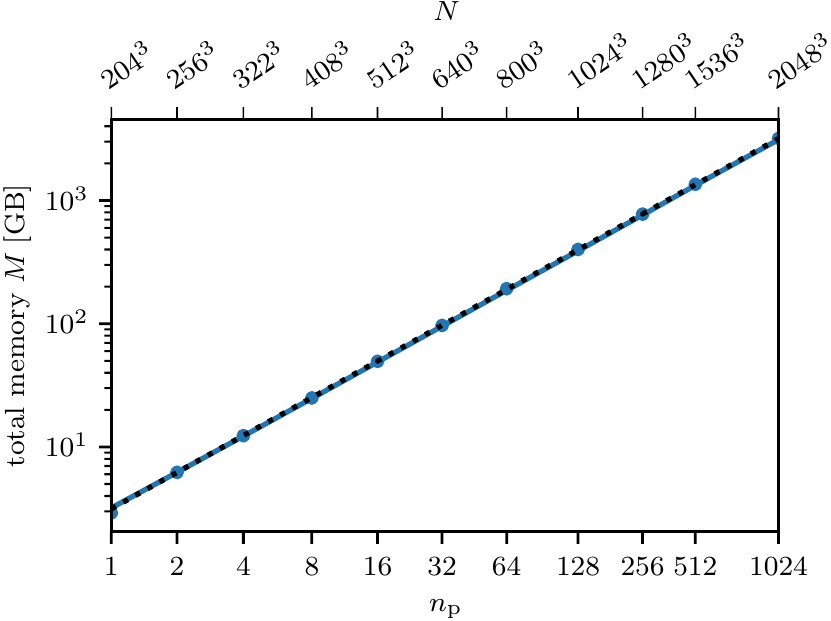}
\caption{Memory scalability for \CONCEPTONE{} simulations in boxes of size $L_{\text{box}} = \SI[parse-numbers = false]{2\cbrt{N}}{Mpc}/h$ at $z=0$, with the number of particles $N$ and processes $n_{\text{p}}$ varying in lockstep. While the blue points are data, the blue line is the estimate \eqref{eq:mem_usage_final}. The dotted black line is perfect scaling $M\propto N$, with outset in the $n_{\text{p}}=2$ data (the serial case $n_{\text{p}}=1$ is not representative due to a lack of communication buffers).
}
\label{fig:mem}
\end{figure}

Factoring in communication buffers (used for sending and receiving particles between processes) and ghost layers, the memory consumption further depends on the number of processes $n_{\text{p}}$. Finally, a constant memory term arises e.g.\ from having the code itself along with libraries loaded into memory. In total, a good memory estimate for \CONCEPTONE{} comes out as
\begin{equation}
	M \approx (0.119 + 0.144\, n_{\text{p}} + \num{3.46e-7}N)\,\si{GB}\,,  \label{eq:mem_usage_final}
\end{equation}
where our standard choice~\eqref{eq:phi_gridsize_preference} has been used to eliminate $n_{\varphi}^3$ in favour of $N$.

Figure~\ref{fig:mem} demonstrates the validity of the memory estimate \eqref{eq:mem_usage_final} for fixed particle resolution. We see that \CONCEPTONE{} follows this estimate nicely, and that the term proportional to $N$ dominates for typical setups, leading to perfect scaling $M\propto N$. Higher particle resolutions will come at a somewhat higher proportionality factor, though with this scaling retained.

Some arrays/buffers within \CONCEPTONE{} have a fixed (maximum) size, for which data too large to fit will be handled in chunks. This is the case for e.g.\ the communication buffers. Other arrays/buffers are free to expand indefinitely, as is the case for e.g.\ the local particle storage within each process. While Figure~\ref{fig:mem} shows the total memory consumption at $z=0$, this number is really varying (almost always monotonically growing) throughout the simulation. Tests show that --- at least for reasonable setups --- this memory growth over time is limited to a few percent.

The memory usage \eqref{eq:mem_usage} is similar to what is reported for \GADGETTWO{} in \citet{gadget2_userguide}, namely $\sim 110$ bytes per particle and 24--32 bytes per \PTHREEM{} grid cell, though this can be halved if using single-precision. While not feasible with \CONCEPTONE{} due to vastly increased computation time, \GADGET{} may be run with a smaller \PTHREEM{} grid than our standard choice \eqref{eq:phi_gridsize_preference}, significantly reducing the memory requirement. As lowering $n_{\varphi}$ shifts more of the computational burden onto the short-range computation, Figure~\ref{fig:box} demonstrates this difference between the two codes nicely, with higher resolution corresponding to more expensive short-range computations and thus smaller $n_{\varphi}$. We note that decreasing $n_{\varphi}$ from $2\cbrt{N}$ to e.g.\ $1\cbrt{N}$ does make \GADGET{} significantly slower as well, but by an acceptable amount in the case of limited memory resources.

In practice, the availability of memory resources is rarely a limiting factor for typical $N$-body simulations, with modern HPC CPUs each having access to hundreds of GB of RAM. With the total memory of simulations scaling as $M\propto N$ and the total computation time as (at best) $\propto N\log N$, the availability of memory will only become less of a problem in the future, assuming similar advances in computational throughput and memory technology.

\subsection{Internal data structures}

\begin{figure*}
\includegraphics[width=\textwidth]{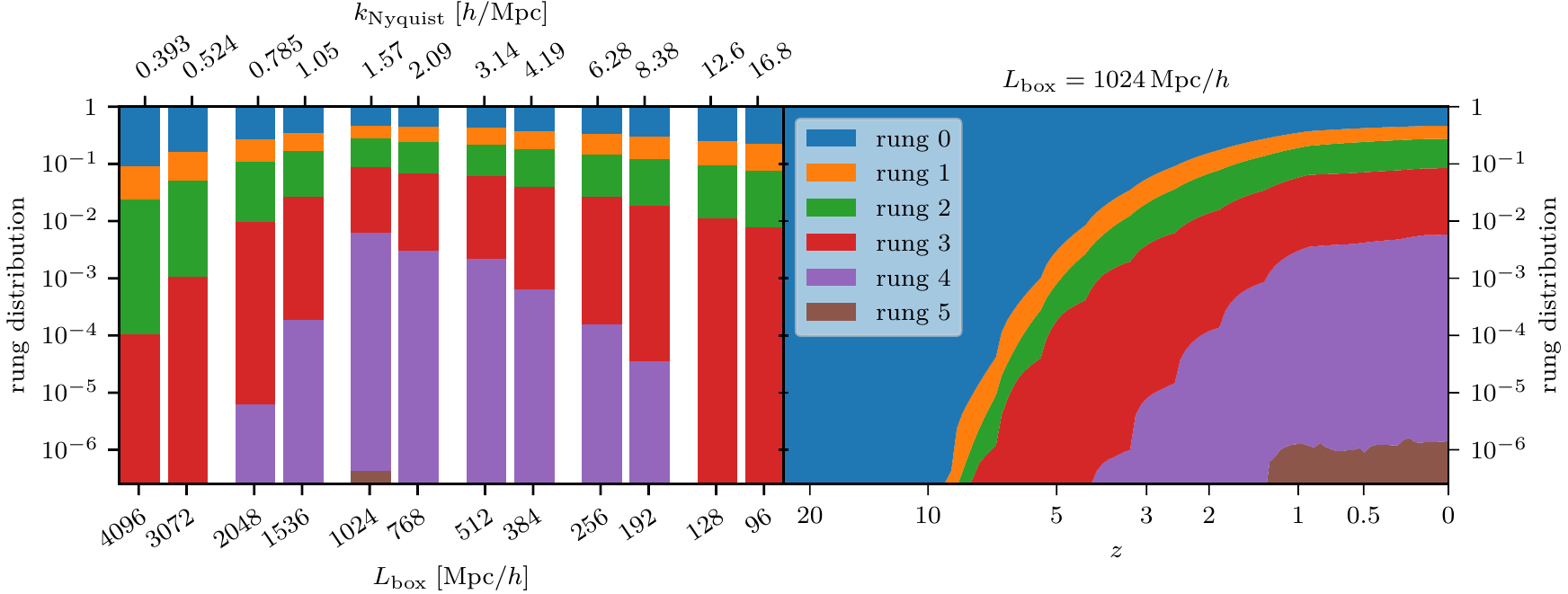}
\caption{Distribution of particles across rungs in \CONCEPTONE{} simulations with $N=512^3$ particles. The left panel shows stacked bar charts of the rung distribution at $z=0$ for simulations with different box sizes $L_{\text{box}}$, or equivalently different particle resolutions $k_{\text{Nyquist}} = \cbrt{N}/2 \times 2\PI/L_{\text{box}} = 512\PI/L_{\text{box}}$. In all cases rung 0 is the most populous one, with the particle count within each rung rapidly declining for the higher rungs. Rungs 4 and 5 are only populated at intermediate box sizes, with particles within simulations in very large or very small boxes only occupying rungs 0--3. The right panel shows the temporal evolution of the rung population for the $L_{\text{box}} = \SI{1024}{Mpc}/h$ simulation, averaged over 8 time steps. All particles start on rung 0 and only begin to jump to higher rungs after $z = 10$. The redshift $z$ axis is shown as scaling linearly with the simulation time steps.
}
\label{fig:rung}
\end{figure*}

The \PTHREEM{} method of \CONCEPTONE{} employs both spatial and temporal adaptiveness in the form of dynamic domain-specific subtiling as described in section~\ref{subsubsec:p3m_gravity} and rung-based particle time-stepping as described in section~\ref{subsubsec:adaptive_timestepping}. With the overall code performance showcased in the previous subsections, let us now take a closer look at these dynamic data structures as a function of time and particle resolution.

\subsubsection{Rung population}
The left panel of Figure~\ref{fig:rung} shows the rung population at $z=0$ in simulations of different particle resolution. For very large boxes, only the few --- here 4 --- lowest rungs are populated. Increasing the particle resolution (lowering the box size) leads to migration of particles to higher rungs, slowly draining rung 0 and now populating rungs 4 and 5 as well. This is expected from the larger particle accelerations (see \eqref{eq:rung}) induced by the increased amount of clustering.

For $k_{\text{Nyquist}} \gtrsim 2\, h/\si{Mpc}$ however, the trend reverses and particles jump back down to the lower rungs. We can understand this perhaps surprising find by considering the interplay between rungs $\ell_i$ \eqref{eq:rung} and the global time step size $\Delta t$. From Figure~\ref{fig:timestepsize} we see that we require $L_{\text{box}} \gtrsim 2\SI[parse-numbers = false]{\cbrt{N}}{Mpc}/h$ in order for the \PTHREEM{} limiter not to dictate a lowering of $\Delta t$ near $z=0$. That is, $L_{\text{box}} \sim 2\SI[parse-numbers = false]{\cbrt{N}}{Mpc}/h$ is the smallest box one can choose before the global time step size is decreased as a result, and so this box size has the largest $\Delta t$ in relation to the amount of clustering. As $L_{\text{box}} \sim 2\SI[parse-numbers = false]{\cbrt{N}}{Mpc}/h$ corresponds to $k_{\text{Nyquist}} \sim 1.57\, h/\si{Mpc}$, this exactly matches the observed behaviour of the left panel of Figure~\ref{fig:rung}.

The right panel of Figure~\ref{fig:rung} shows the time evolution for the simulation with $L_{\text{box}} = 2\SI[parse-numbers = false]{\cbrt{N}}{Mpc}/h$ or equivalently $k_{\text{Nyquist}} = 1.57\, h/\si{Mpc}$. All particles start at rung 0 and stay there until a little after $z = 10$, after which rungs 1--3 are quickly populated, followed by rung 4 at $z\sim 4$ and finally rung 5 at $z\sim 1.5$, though with each higher rung occupying much fewer particles than the ones below. That non-linearity commence at around $z\sim 10$ is consistent with the sudden increase in short-range computation time seen in the lower panel of Figures~\ref{fig:weak}~and~\ref{fig:strong}, which we now understand as arising from an increase in kick operations due to additional rungs being populated.

The bulby look of the evolution of each rung count on the right panel of Figure~\ref{fig:rung} reflects the time step cycle of 8 steps, as described in section~\ref{subsubsec:global_time_step_size}. At the end of each cycle, the global time step $\Delta t$ is allowed to increase, prompting higher rungs as specified by \eqref{eq:rung}. With all particles moving to their newly assigned rung before the next cycle begins, this results in steep increases to the count of rungs $\ell > 0$.

At $z \sim 1$ a qualitative change in behaviour is seen for the rung population of the right panel of Figure~\ref{fig:rung}, where instead of migrating to higher rungs with time, the particles all more or less stay on their given rung throughout the rest of the simulation. Once more we can understand this from Figure~\ref{fig:timestepsize}, where $z\sim 1 \Rightarrow a \sim 0.5$ is where the (upper/rightmost, $L_{\text{box}} = \text{\textonehalf}\SI[parse-numbers = false]{\cbrt{N}}{Mpc}/h$) \PTHREEM{} limiter begins to dominate, no longer allowing the global time step size $\Delta t$ to drastically increase with each finished time step cycle. As the value of the \PTHREEM{} limiter is determined by the root mean square velocity of the particle distribution itself (see section~\ref{subsubsec:global_time_step_size}), the global time step size $\Delta t$ is now evolved in sync with the velocity distribution of the particles, hence why the particles now remain satisfied occupying the same rung for the rest of the simulation. Looking again at the lower panel of Figures~\ref{fig:weak}~or~\ref{fig:strong}, this change in behaviour is once again seen in the short-range computation times, as the slopes suddenly decrease at $z \sim 1$. As the $z$ axes are all shown as scaling linearly with time steps (as opposed to e.g.\ $z$ itself, $a$ or $t$), this slope is proportional to the increase in computation time from one step (or cycle) to the next.

\begin{figure*}
\includegraphics[width=\textwidth]{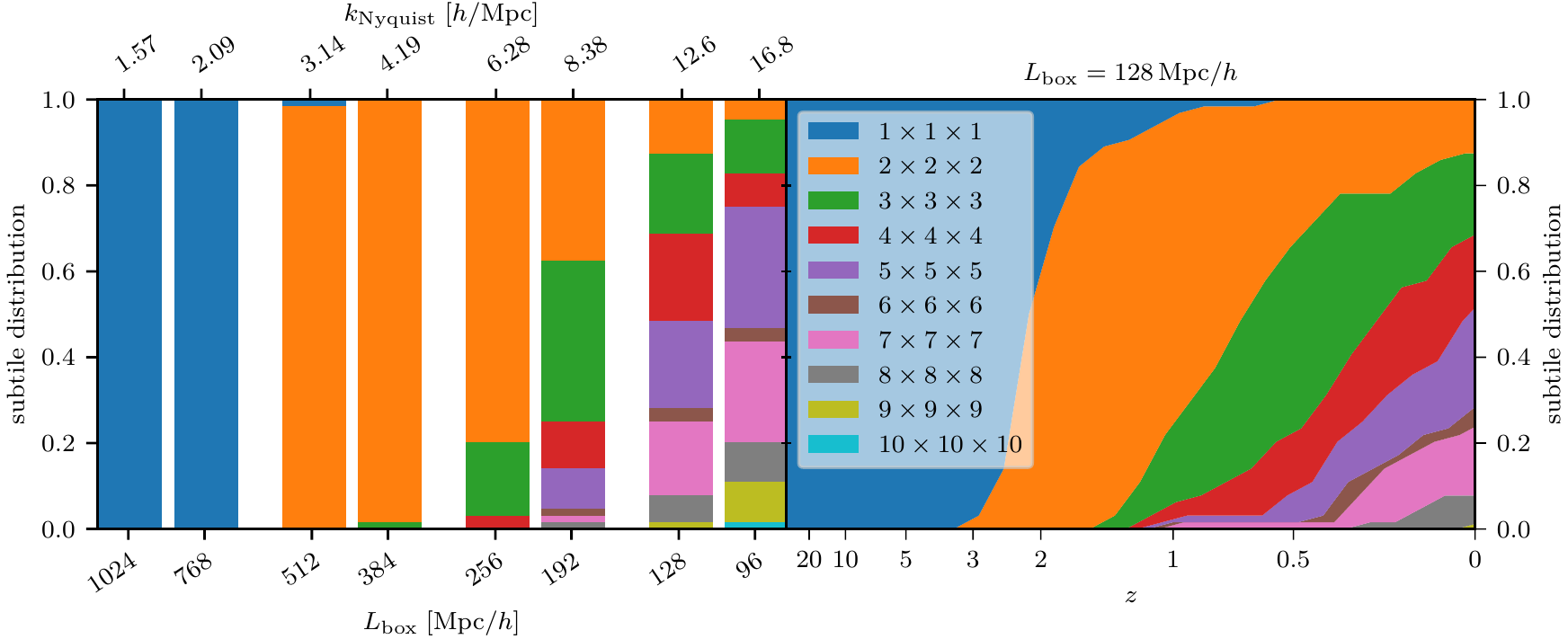}
\caption{Distribution of automatically chosen subtile decompositions across domains for simulations with $N = 512^3$ particles run on $n_{\text{p}} = 64$ processes. The left panel shows stacked bar charts of the distribution of subtile decompositions at $z = 0$ for simulations with different box sizes $L_{\text{box}}$, or equivalently different particle resolutions $k_{\text{Nyquist}} = \cbrt{N}/2 \times 2\PI/L_{\text{box}} = 512\PI/L_{\text{box}}$. Within the two simulations of lowest particle resolution, all domains employ a subtile decomposition of $1\times 1\times 1$. Within each of the next two simulations with slightly higher resolution, all but one domain ends up using the $2\times 2\times 2$ subtile decomposition. Simulations of still higher resolution show diverse distributions of subtile decompositions. The right panel shows the temporal evolution of the subtile decomposition for the $L_{\text{box}} = \SI{128}{Mpc}/h$ simulation. By default, the domains undergo (possible) subtile refinement at every 16\textsuperscript{th} time step --- corresponding to two time step cycles --- and so the data shown has been averaged over 16 time steps. All domains start out using the trivial $1\times 1\times 1$ subtile decomposition and only begin transitioning to $2\times 2\times 2$ at $z\sim 3$, followed by finer decompositions at $z\sim 1.5$. The redshift $z$ axis is shown as scaling linearly with the simulation time steps.
}
\label{fig:subtile}
\end{figure*}

\subsubsection{Subtile decomposition}

The left panel of Figure~\ref{fig:subtile} shows the distribution of automatically chosen subtile decompositions across domains at $z=0$ for simulations with $N = 512^3$ particles run on $n_{\text{p}} = 64$ processes, for different particle resolutions. For low particle resolutions $k_{\text{Nyquist}} \lesssim 2\, h/\si{Mpc}$, every domain employs the trivial $1\times 1\times 1$ subtile decomposition, corresponding to not subdividing tiles into subtiles at all. Proper subtile decompositions are in use by all domains for simulations with $k_{\text{Nyquist}} \gtrsim 4\, h/\si{Mpc}$, with very fine subtile decompositions quickly following for still higher resolutions. In the case of high clustering it is clear that the automatic subtile refinement in \CONCEPTONE{} prefers not just a fine global subtile decomposition, but a subtile decomposition with great spatial variation.

The right panel of Figure~\ref{fig:subtile} shows the time evolution of the distribution of subtile decompositions within the $L_{\text{box}} = \SI{128}{Mpc}/h$ ($k_{\text{Nyquist}} = 12.6\, h/\si{Mpc}$) simulation. We see that even with this high resolution, all domains utilise the trivial $1\times 1\times 1$ decomposition until $z \sim 3$. Slightly further in the time evolution, the majority of the domains switch to the $2\times 2\times 2$ decomposition. For $z \lesssim 1$ a more inhomogeneous distribution arise, demonstrating varying amounts of non-linearity within the domains. A significant fraction of the domains evolve very finely subdivided decompositions at $z \lesssim 0.5$.

\begin{figure*}
\includegraphics[width=\textwidth]{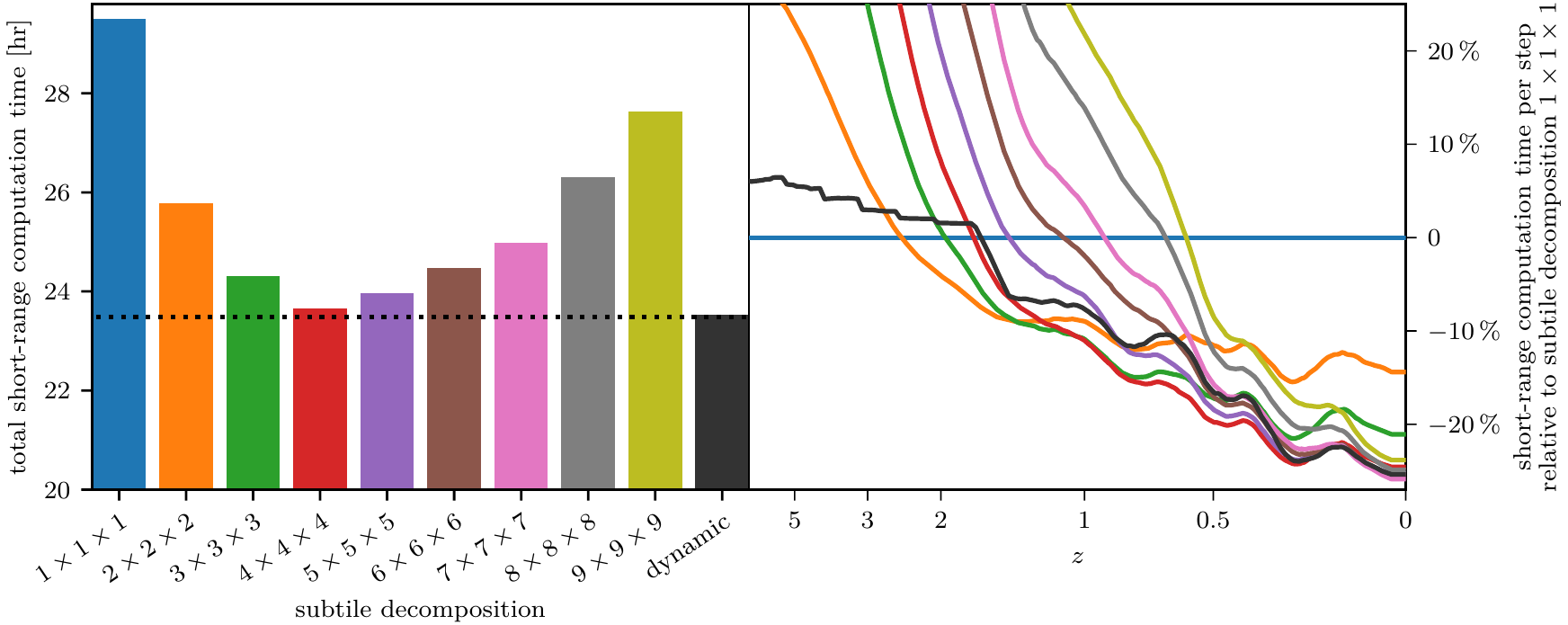}
\caption{Wall-clock computation times for the short-range force in simulations with $N=512^3$ particles in a box of size $L_{\text{box}} = \SI{192}{Mpc}/h$, run on $n_{\text{p}} = 64$ processes. The left panel shows the cumulative short-range computation time throughout simulations using different fixed subtile decompositions, together with that of a simulation with automatic/dynamic subtile refinement. A clear minimum in computation time is found from using a fixed subtile decomposition of $4\times 4\times 4$. Using automatic/dynamic subtile refinement is slightly faster still, as the horizontal dotted line helps to show. The right panel shows the evolution of the single-step short-range computation time for the different choices of subtile decomposition, relative to that of the trivial $1\times 1\times 1$ decomposition, averaged over 16 time steps. The colours of the lines match those of the bars on the left panel. While the trivial tiling is initially superior, it is eventually overtaken by the higher fixed decompositions and the dynamic decomposition. The redshift $z$ axis is shown as scaling linearly with the simulation time steps.
}
\label{fig:subtile_refinement}
\end{figure*}

While \CONCEPTONE{} does support subtile decompositions with a different number of subdivisions along each dimension (see Figure~\ref{fig:geometry}), our cubic choice of $n_{\text{p}} = 64 = 4^3$ leads to all automatically chosen subtile decompositions being cubic as well. This happens because the domains become perfect cubes, which in turn leads to cubic tiles. When subdividing the tiles into subtiles, it is generally done in a manner so that the subtiles end up as cubic as possible, which here leads to all three dimensions being subdivided in lockstep, and so only cubic subtile decompositions appear in Figure~\ref{fig:subtile}.

The evolution of the subtile decompositions does not depend directly on the particle dynamics, but only indirectly on the non-linear clustering through the computation time of the short-range force, as described towards the end of section~\ref{subsubsec:p3m_gravity}. It is thus hard to quantitatively relate Figure~\ref{fig:subtile} to what we see in previous figures, except to say that the non-linearity definitely grows monotonically with both time and particle resolution, within the box as a whole and within each domain separately.

We have yet to show how much of a performance improvement the subtiles actually bring, and whether letting \CONCEPTONE{} automatically and dynamically refine the subtiling actually leads to smaller computation times than what can be achieved by running with fixed subtile decomposition. To this end, the left panel of Figure~\ref{fig:subtile_refinement} shows the total short-range computation time accumulated throughout simulations of $k_{\text{Nyquist}} = 12.6\, h/\si{Mpc}$, as also used for Figure~\ref{fig:subtile}. The figure shows computation times for simulations with fixed, cubic subtile decompositions from $1\times 1\times 1$ to $9\times 9\times 9$, with a minimum at $4\times 4\times 4$ and clear trends of increased computation times when moving away from this choice in either direction. The slowest of all is $1\times 1\times 1$, being a significant\footnote{As seen from Figures~\ref{fig:weak}~and~\ref{fig:strong}, the cumulated short-range computation time in fact comprises the vast majority of the total simulation time. For the high particle resolution of Figure~\ref{fig:subtile_refinement}, any reduction in short-range computation time directly translates into the same reduction in total computation time, to within a few percent.} $\SI{20}{\percent}$ slower than $4\times 4 \times 4$. It is interesting that even fixed $9\times 9\times 9$ is superior to fixed $1\times 1\times 1$, considering that the trivial $1\times 1\times 1$ is actually preferred throughout much of the simulation time span, according to the dynamic subtile refinement of the right panel of Figure~\ref{fig:subtile}, with $9\times 9\times 9$ only being used within a single domain towards the end. This shows that --- at least for simulations of high resolution --- dividing tiles into subtiles is a rather robust optimisation.

The standard choice of automatic and dynamic subtile refinement is also shown on the left panel of Figure~\ref{fig:subtile_refinement}, where it just manages to outperform $4\times 4 \times 4$, though only by $\SI{0.5}{\percent}$. This is typical for lower particle resolutions as well, and so the dynamic subtiling feature is really not significantly better than just fixing the subtile decomposition to the optimal value. However, exactly which decomposition to use if running with a fixed decomposition is not easy to determine, as it depends not only on particle resolution and cosmology, but also the shape of the tiles, which in turn depends on the number of processes $n_{\text{p}}$ (through the domain decomposition) and the short-range cut-off scale $x_{\text{r}}$ (as described in section~\ref{subsubsec:p3m_gravity}). Regarding the computation time, running with automatic subtile refinement is then more or less equivalent to always running with the best possible fixed subtile decomposition.

The right panel of Figure~\ref{fig:subtile_refinement} shows how the different choices of subtile decomposition performs over the course of the simulation. Initially, until $z\sim 2.5$, the trivial $1\times 1\times 1$ decomposition is fastest, as expected for this case of low clustering. Interestingly, though the dynamic decomposition also uses $1\times 1\times 1$ within all domains at these early times (see Figure~\ref{fig:subtile}), it is significantly slower. This is due to wasted effort trying out the higher decomposition of $2\times 2\times 2$ at every 16\textsuperscript{th} time step, only to discard it again once measurements reveal it to be inferior to $1\times 1\times 1$. In fact, the dynamic decomposition is very rarely ever the fastest instantaneous choice due to this reason. Yet, when integrated\footnote{As the $z$ axis of the right panel of Figure~\ref{fig:subtile_refinement} scales linearly with the time steps, this integral is in fact exactly proportional to the (signed) visual area between the various lines and the horizontal $1\times 1\times 1$ line.} over the whole simulation time span, it comes out as the most performant.

\section{Discussion and conclusions}\label{sec:conclusion}
In this paper we have presented the new massively parallel cosmological structure formation simulation code \CONCEPTONE{}. This is the first of its kind written in Python, yet it achieves performance comparable to existing state of the art codes, such as \GADGETTWOFOUR{}.

The code contains an efficient \PTHREEM{} gravity solver, achieving excellent performance through a combination of a large-scale potential solver and direct summation over short distances. Unlike e.g.\ tree codes, the short-range force in \CONCEPTONE{} is basically exact. This is implemented using a novel subtiling scheme, leading to significant speed-ups by lowering the number of false positive particle pairs. The \PTHREEM{} gravity is coupled with individual and adaptive particle time-stepping, allowing for efficient time integration with high temporal resolution. These numerical methods are presented and built up from first principles. As such, section~\ref{sec:numerical_methods} on its own serves as a great reference for the key numerical methods employed by cosmological $N$-body codes.

We have run a large number of scaling tests of the code. We find that \CONCEPTONE{} exhibits excellent scaling behaviour --- weak as well as strong --- up to at least a thousand CPU cores, though we expect good scaling up to significantly larger core counts, for reasonable workloads.

We have in fact performed a few larger simulations as well, the largest being $N = 4028^3$ particles run on $n_{\text{p}} = 4028$ CPU cores. While the good scaling behaviour seems to be more or less retained for such large simulations, the FFTs inevitably begin to dominate the computation time, a trend that can be seen from the upper left panels of Figures~\ref{fig:weak}~and~\ref{fig:strong}. As the slab-decomposition of the \PTHREEMPAREN{} grid inherited from FFTW forces $n_{\text{p}} \leq n_{\varphi} \sim \cbrt{N}$, this poses a problem for still larger simulations.

Many of the observed details regarding the performance throughout the simulation time span are successfully explained by considering the implemented time-stepping scheme and internal data structures. We find the total memory footprint of \CONCEPTONE{} to be proportional to the number of particles $N$, as expected. While the absolute memory usage is reasonable, \CONCEPTONE{} is not written to be especially `lean'.

We have gauged the accuracy of \CONCEPTONE{} by comparing its results to those of \GADGETTWOFOUR{}, at the power spectrum level. Here we find truly extraordinary agreement between \CONCEPTONE{} and \GADGET{}, at the sub-percent level for all times and most scales. For all but the smallest scales, the agreement is even below one per mille, with especially good agreement observed for the new and improved \GADGETFOUR{} running with a high multipole expansion order. This is remarkable, as \CONCEPT{} to a large extent has been modelled on and tested against \GADGETTWO{} throughout its development.

We have tested the absolute performance of the code by comparing computation times of full \CONCEPTONE{} simulations to those of equivalent \GADGETTWO{} simulations. Here we find \CONCEPTONE{} to be several times faster for simulations of low to medium resolutions ($k_{\text{Nyquist}} \lesssim 5\,\si{Mpc}/h$), while the opposite is true for higher-resolution simulations. We give two reasons for the superior speed of \CONCEPTONE{} at large scales. First, the (sub)tiled short-range implementation of \PTHREEM{} within \CONCEPTONE{} is extremely efficient at low clustering. Second, the early time-stepping within \CONCEPTONE{} is more aggressive (though benign) than it is within \GADGET{}. For simulations of very high resolution, the direct short-range summation within \CONCEPTONE{} becomes prohibitively expensive, due in part to intrinsics of the algorithm, but also due to the current lack of load balancing. It should be noted that the gravitational computation within \CONCEPTONE{} is inherently more accurate than that of \GADGET{} due to \CONCEPTONE{} not using a tree, and so this performance comparison is really done with the two codes running at unequal accuracies. Increasing the accuracy of the tree in \GADGETTWO{}, we find both longer computation times and better agreement with the results of \CONCEPTONE{}. Likewise, we generally find significantly better small-scale agreement between \CONCEPTONE{} and \GADGETFOUR{} than between \CONCEPTONE{} and \GADGETTWO{}, and that the \GADGETFOUR{} simulations are 2--3 times as expensive to perform as their \GADGETTWO{} counterparts.

The focus of this paper has been on the last few year's overhaul of the \CONCEPT{} code, culminating in version 1.0. The primary new capabilities compared to earlier versions are those of short-range forces and adaptive time-stepping, the numerical methods of which have been described. The many older features are however retained in \CONCEPTONE{}, which makes the code extremely well-suited for simulations of cosmological structure formation, in particular in large boxes. The \CONCEPTONE{} code works in $N$-body gauge and can produce output which is fully compatible with general relativistic perturbation theory, including both relativistic corrections to the particle equations of motion as well as perturbations from linearly clustered species, such as photons and neutrinos. This also allows \CONCEPTONE{} to run fully self-consistent simulations of a variety of non-standard cosmological models, such as decaying dark matter and dynamical dark energy. A non-linear fluid solver is also built-in, allowing \CONCEPTONE{} to produce accurate estimates of the distribution of massive (but light) neutrinos without the particle noise seen in particle-based neutrino simulations. On top of all this, a slew of different numerical schemes and other features are also implemented, many of which are new to the 1.0 release. We briefly list these in appendix~\ref{subsec:additional_features}.

Finally, there are a number of performance limitations in the current version 1.0 of \CONCEPT{}, which will be addressed in future releases. First, for strongly clustered systems the static domain decomposition employed leads to substantial load imbalance, significantly reducing the performance compared to the theoretical optimum. We plan to address this issue by utilising a structure similar to the pseudo-Hilbert curve used in \GADGET{}, but at the tile level rather than the particle level.
Second, \CONCEPTONE{} uses the standard slab-decomposition of FFTW, which limits the number of CPU cores participating in the long-range force computation to be less than or equal to the grid size $n_{\varphi}$ of the \PTHREEMPAREN{} grid. This can be remedied by switching to a pencil-decomposition, as used in e.g.\ \GADGETFOUR{}.

\section*{Acknowledgements}
We wish to thank Volker Springel for valuable discussions, in particular on the code comparisons between \CONCEPTONE{} and \GADGETTWOFOUR{}.
We are thankful to Tiago Castro for pointing out several bugs and shortcomings of \CONCEPT{} prior to the 1.0 release.
We thank Joachim Harnois-D{\'e}raps for many insightful comments on the manuscript.
We acknowledge computing resources from the Centre for Scientific Computing Aarhus (CSCAA).
J.D.\ and T.T.\ was supported by a research grant (29337) from VILLUM FONDEN.

\section*{Data availability}
The \CONCEPTONE{} code discussed in this paper is openly released at \href{https://github.com/jmd-dk/concept}{github.com/jmd-dk/concept}\,. Data and scripts used for generating all figures within this paper --- as well as simulation parameter files for recreating the simulations producing this data --- are made publicly available at \href{https://github.com/AarhusCosmology/concept1.0-data}{github.com/AarhusCosmology/concept1.0-data}\,.


\bibliographystyle{mnras}
\bibliography{bibliography}


\appendix

\section{Other code aspects}\label{sec:other}
The purpose of this appendix is to provide brief overviews of secondary aspects of \CONCEPTONE{}, specifically the many built-in features besides particle dynamics, as well as the unusual though modern software framework in which \CONCEPTONE{} operates.

\subsection{Additional features}\label{subsec:additional_features}
Though this paper focuses on the core $N$-body functionality of \PTHREEM{} and adaptive time-stepping, \CONCEPTONE{} in fact contains a lot of additional features. Here some of these are briefly listed.
\begin{description}
	\item Multiple possible non-linear components using either a particle or fluid representation; cold dark matter, decaying cold dark matter \citep{concept_dcdm}, massive neutrinos \citep{nuconcept}. Various parameters are tuneable at a per component basis.
	\item Linear components, allowing for simulations consistent with general relativistic perturbation theory; photons, massive/massless neutrinos \citep{concept_linnu}, dynamical dark energy \citep{concept_de}, dark radiation \citep{concept_dcdm}.
	\item On-the-fly initial condition generation of all implemented species, either using standard Gaussian noise or the `paired-and-fixed' technique of \citet{pairedfixed}.
	\item Complete integration of \CLASS{} \citep{class}, used to obtain background values such as $a(t)$ and linear perturbations for initial conditions and linear components. All perturbations are transformed to $N$-body gauge so that simulation results can be interpreted in a relativistic setting \citep{concept_linnu}.
	\item Output: Power spectra (data, image), 2D renders (data, image, terminal visualisation), 3D renders (image), snapshots. The \CONCEPTONE{} code implements its own snapshot format (HDF5) --- capable of storing particle and fluid components --- as well as the full\footnote{Including reading and writing snapshots of \texttt{SnapFormat} 1 and 2, multi-file snapshots, multiple particle types, 32- or 64-bit particle data and IDs, and more.} specification of the well-known binary \GADGET{} format \citep{gadget2}.
	\item Grids used for \PTHREEMPAREN{} \eqref{eq:p3m_gravity}, power spectra and 2D renders may use any of the implemented interpolations \eqref{eq:W_NGP,eq:W_CIC,eq:W_TSC,eq:W_PCS} with optional deconvolution \eqref{eq:weight_fourier} as well as optional interlacing\footnote{By default, deconvolutions are always on, while interlacing is enabled for power spectra but not for \PTHREEMPAREN{}.} \citep{HockneyEastwood}. The grid size of each component is independent, with collective grids computed by adding up (properly shifted) Fourier values, used when e.g.\ several components contribute to the \PTHREEMPAREN{} grid or when computing combined auto-spectra of multiple components.
	\item The \PTHREEMPAREN{} grid force may be obtained from the potential either by real-space differentiation \eqref{eq:symmetric_diff_2,eq:symmetric_diff_4,eq:symmetric_diff_6,eq:symmetric_diff_8} or using Fourier-space differentiation.
	\item A robust and easy-to-use autosave mechanism is available, periodically saving the state of the simulation to disk.
	\item By default, simulation results are not exactly deterministic (e.g.\ due to the non-associativity of floating-point addition), though this can be toggled through parameters.
	\item Various auxiliary \emph{utilities} are included alongside the main code, which provide functionality outside of running simulations, such as computing power spectra directly from snapshots.
	\item All user interaction happens through a script with discoverable command-line options, which handles building (on modification) of the code, job execution and even submission via Slurm/TORQUE/PBS.
	\item Complete and flexible installation script for automated installation of \CONCEPTONE{} --- along with all of its dependencies --- with no special permissions required. Successfully tested on dozens of Linux clusters, servers and laptops.
	\item Docker images of \CONCEPTONE{} are freely available on Docker Hub\footnote{\href{https://hub.docker.com/r/jmddk/concept}{hub.docker.com/r/jmddk/concept}}, convenient for quickly trying out the code.
	\item Large suite of integration tests for continuous code validation. As the installation depends on online resources, the installation along with the entire test suite is automatically tested periodically on GitHub, with the latest result publicly visible.
	\item Thorough documentation\footnote{\href{https://jmd-dk.github.io/concept}{jmd-dk.github.io/concept}} --- including an expansive tutorial --- of how to use the code is publicly released together with the source.
\end{description}

\subsection{Code language and build process}\label{subsec:language}
Though no knowledge of the internals of \CONCEPT{} is needed in order to make use of the code, we here give a brief overview, as the technology employed is rather novel.

Today, most scientific code gets written using higher-level languages, probably mainly due to the rapid development these languages and their ecosystems allow for. These languages are typically dynamical and interpreted, which comes at a performance penalty. High-performance simulation codes are thus still primarily written in low-level languages such as Fortran, C or \CPP{}. While allowing for performant code where needed, this further forces the lower level aspects upon the rest of the code base, with no performance benefits. This generally makes the code harder to read and extend, especially for the many scientists not fluent in such languages.

The most prevalent high-level language used for scientific computing in the current era is arguably Python, which is also the language chosen for \CONCEPT{}. While performance to some extent is obtainable through the use of numerical libraries such as NumPy \citep{numpy} and FFTW \citep{fftw}, this is not enough to compete with high-performance low-level codes such as \GADGET{}. To this end, \CONCEPT{} makes heavy use of Cython \citep{cython}, which translates Python code to equivalent C code, which must then be compiled as any other C program. By further specifying the types of key variables, the translated result can be made as good as hand-written C.

While Cython does allow for seamless mixing of dynamic Python code and typed ``C-like'' Python code, some of its low-level features (e.g.\ access to raw pointers) require use of syntax that breaks Python compatibility, meaning the code now \emph{only} runs after transpilation to C. As rapid development and debugging relies heavily on the code being executable as a pure Python script, \CONCEPT{} effectively implements its own language on top of Cython, with new Python-compatible syntax for these missing functionalities.

While the raw \CONCEPT{} source code may then be executed directly in Python, it can alternatively (and preferably) be built by first transpiling it to valid Cython code\footnote{While with standard Cython one has to further write a header file per code file (as in C), we have automated this task as part of the built-in transpiler. Thus the source code consists solely of the bare Python files, with everything else generated from this.} using a custom built-in transpiler, after which the code is further transpiled to C using the Cython transpiler, and then finally compiled to machine code using a C compiler. This entire build process is of course automated.

Besides serving as a bridge between Python and low-level Cython, the custom transpiler further enables quite a few performance enhancements through direct source code transformations. These include early run-time or even compile-time expression evaluation, loop unswitching and iterator inlining. Oftentimes these are optimisations which cannot be applied by the C compiler itself and which are not easily or conveniently expressible in low-level languages such as C.


\bsp  
\label{lastpage}
\end{document}